\documentclass[twocolumn]{jpsj2} 
\title{Non-Fermi liquid behavior in the magnetotransport of  Ce$M$In$_5$ ($M$: Co and Rh): Striking similarity between quasi two-dimensional heavy fermion and high-$T_c$ cuprates}

\author{Y. \textsc{Nakajima}$^{1,2}$, H. \textsc{Shishido}$^{1}$, H. \textsc{Nakai}$^{1}$, T. \textsc{Shibauchi}$^{1}$,  K. \textsc{Behnia}$^{3}$, K. \textsc{Izawa}$^{2,4}$, M. \textsc{Hedo}$^{2}$, Y. \textsc{Uwatoko}$^{2}$, T. \textsc{Matsumoto}$^{5}$,  R. \textsc{Settai}$^{6}$,  Y. \textsc{Onuki}$^{6}$, H. \textsc{Kontani}$^{7}$, and Y. \textsc{Matsuda}$^{1,2}$}

\inst{
$^{1}$Department of Physiscs, Kyoto University, Kyoto 606-8502, Japan\\
$^{2}$Institute for Solid State Physics, University of Tokyo, Kashiwanoha, Kashiwa, Chiba 277-8581, Japan\\
$^{3}$Laboratoire de physique quantique (CNRS), ESPCI, 10 Rue Vauquelin, 75005 Paris, France\\
$^{4}$CEA-Grenoble, 38054 Grenoble Cedex 9,  France\\
$^{5}$National Institute of Material Science, Sakura, Tsukuba, Ibaraki 305-0003, Japan\\
$^{6}$Graduate School of Science, Osaka University, Toyonaka, Osaka 560-0043, Japan\\
$^{7}$Department of Physics, Nagoya University, Furo-cho, Chikusa-ku, Nagoya 464-8602, Japan\\
}

\abst{We present a systematic study of the dc-resistivity, Hall effect, and magnetoresistance in the normal state of quasi two-dimensional (2D) heavy fermion superconductors Ce$M$In$_5$ ($M$: Rh and Co) under pressure.  Here the electronic system evolves with pressure from an antiferromagnetic (AF) metal, through a highly unconventional non-Fermi liquid, and finally into a Fermi-liquid state. The novelty of these materials is best highlighted when compared with La$M$In$_5$, a system with similar electronic structures, which shows a nearly temperature independent Hall coefficient and a magnetoresisitance which is well described by the classical Kohler rule.  In sharp contrast, in Ce$M$In$_5$, the amplitude of the Hall coefficient increases dramatically with decreasing temperature, reaching at low temperatures a value significantly larger than $1/ne$, where $n$ is the carrier number. Furthermore, the magnetoresistance is characterized by $T$- and $H$-dependence which clearly violate Kohler's rule.  We found that the cotangent of the Hall angle $\cot \Theta_H$ varies as $T^2$, and the magnetoresistance is well scaled by the Hall angle as $\Delta \rho_{xx}/\rho_{xx}\propto \tan^2\Theta_H$. These non-Fermi liquid properties in the electron transport are remarkably pronounced  when the AF fluctuations are enhanced in the vicinity of the quantum critical point.  We lay particular emphasis on the striking resemblance of these anomalous magnetotransport with those of the high-$T_c$ cuprates.  We argue that features commonly observed in quasi 2D heavy fermion and cuprates very likely capture universal features of strongly correlated electron systems in the presence of strong AF fluctuations, holding the promise of bridging our understanding of  heavy fermion systems and high-$T_c$ cuprates. }

\kword{Hall effect, magnetoresistance, Fermi liquid, heavy fermion, quantum critical point}

\begin{document}

\maketitle

\section{Introduction}

One of the most interesting and puzzling issues in the research of strongly correlated electron systems is anomalous electron transport phenomena.   The dc-resistivity, Hall coefficient, and magnetoresistance are  fundamental transport property parameters and it is well established that in conventional metals these transport parameters are well described by Landau's Fermi liquid theory.    In the Fermi liquid state, the dc resistivity,  Hall coefficient, and magnetoresistance exhibit chracteristic $T$- and $H$-dependence. The dc resistivity $\rho_{xx}$ due to electron$-$electron scattering depends on $T$ as
\begin{equation}
\rho_{xx}\propto T^2.
\end{equation}
The Hall coefficient $R_H$ signifies the Fermi surface topology and carrier density; in the simple case it is given as 
\begin{equation}
R_H=\frac{1}{ne},
\end{equation}
where $n$ is the carrier number, and is independent of temperature \cite{hall}.  The magnetoresistance stems from bending of the electron trajectory by the Lorentz force.   The magnetoresistance can be used to probe deviations of the Fermi surface from sphericity, providing unique information on how the electron mean free path varies around the Fermi surface. The magnetoresisitance $\Delta\rho_{xx}/\rho_{xx}$ obeys  Kohler's rule, 
\begin{equation}
\frac{\Delta\rho_{xx}}{\rho_{xx}}=F\left(\frac{H}{\rho_{xx}}\right),
\end{equation}
where $\Delta\rho_{xx}\equiv \rho_{xx}(H)-\rho_{xx}(0)$ and  $F(x)$ is a function depending on the details of the electronic structure \cite{pippard}.

Within the last decade, however,  it has been found that in strongly correlated materials, including heavy fermion intermetallics, organics, and high-$T_c$ cuprates, these transport parameters display striking deviations from  Fermi liquid behavior, especially in the presence of the strong magnetic fluctuations.\cite{moriya,coleman,sachdev}.  In each of these materials, chemical doping,  pressure, or magnetic field can tune the quantum and thermal fluctuations at zero and finite temperatures.   It is generally believed that the strong magnetic fluctuations seriously modify the quasiparticle masses and scattering cross section of a Fermi liquid.   As a result, these systems develop a new excitation structure and often display peculiar electron transport properties over a wide temperature range.  In addition, some of these metals display  superconductivity.  It is also generally believed that the strong electron correlation leads to a notable many-body effect and often gives rise to unconventional superconducting pair states with angular momentum greater than zero \cite{sigrist,thalmeier,mineev}.   Therefore  the properties of the normal state, which is the stage of the novel superconductivity, are intimately related to the mechanism of the superconducting pairing interaction.

Among strongly correlated materials, the electron transport properties of high-$T_c$ cuprates have been studied the most extensively.  In the optimally and underdoped regimes, they exhibit all the hallmarks of non-Fermi liquid behavior.   In the optimally doped regime, the resistivity shows a $T$-linear dependence over a wide temperature range from $T_c$ up to very high temperatures:
\begin{equation}
\rho_{xx}\propto T.
\end{equation}
The Hall coefficient $R_H$ exhibits strong,  sometimes peaked,  temperature and doping dependence \cite{ong1,takeda,ando}.  At very high temperatures,  $R_H$ is close to $1/ne$, as expected for conventional  metals.   However,  $R_H$ increases with decreasing temperature and attains a value much larger than $1/ne$ at low temperatures.  It has been pointed out that  the Hall problem can be simplified when analyzed in terms of $\cot \Theta_H$,  where $\Theta_H\equiv  \tan^{-1} (\rho_{xy}/\rho_{xx})$ is the Hall angle and $\rho_{xy}$ is the Hall resistivity \cite{chien,anderson}.  In high-$T_c$ cuprates, the Hall angle approximately varies as
\begin{equation}
\cot \Theta_H \equiv \frac{\rho_{xx}}{\rho_{xy}}\propto T^2. 
\end{equation}
The magnetoresistance does not show the scaling relation given in Eq.~(3), indicating a strong violation of Kohler's rule.  A new scaling relation,
\begin{equation}
\frac{\Delta \rho_{xx}}{\rho_{xx}}\propto \tan ^2\Theta_H, 
\end{equation}
has been proposed to describe the $T$- and $H$-dependence of the magnetoresistance \cite{harris}.  In the underdoped regime, $\rho_{xx}$ and $R_H$ are reduced below the (smaller) pseudogap temperature.    In addition to these electron transport properties,  the thermoelectric effects have been reported to be quite anomalous in the underdoped and optimally doped regimes.  In particular a giant enhancement of the Nernst coefficient $\nu_H$,   three orders of magnitude larger than that expected for conventional metals, has been reported for high-$T_c$ cuprates \cite{nernst}.

Extensive experimental and theoretical studies have been made to establish a picture of the basic transport properties in non-Fermi liquid states.  Two types of theoretical models have been proposed: models based on a Fermi liquid ground state, in which the carriers are electrons and holes, and more exotic models based on a non-Fermi liquid ground state, in which the transport carriers are exotic objects, such as spinons and holons.   In the Fermi liquid approach, an anisotropic scattering mechanism of carriers is invoked \cite{pines,millis,kontani,rosch,hussey}.  The central concept behind this is the ''hot spot", a small region on the Fermi surface where the electron lifetime is unusually short, e.g., due to the electron$-$antiferromagnetic(AF) spin fluctuation scattering.   In the non-Fermi liquid approach,   the charge transport is assumed to be governed by  two separate  scattering times with different temperature dependence associated with  spin$-$charge separation \cite{chien,anderson}.

There appears to be common agreement on the inadequacy of the Bloch$-$Boltzmann approach (i.e., the relaxation time approximation) for describing transport experiments on strongly correlated electron systems \cite{ong2}.  However there is no established explanation for the non-Fermi liquid behavior observed in the basic transport properties.  Therefore the anomalous $T$- and $H$-dependence of the dc resisticty,  Hall effect, and  magnetoresistance remains one of the most important unresolved issues in strongly correlated electron systems; and an important outstanding question related to this issue is {\it whether the anomalous transport properties shown in Eqs.~(4)$-$(6) are specific to high-$T_c$ cuprates or represent universal features in strongly correlated electron systems in the presence of strong AF fluctuations}.

\begin{figure}[t]
\begin{center}
\includegraphics[width=8.5cm]{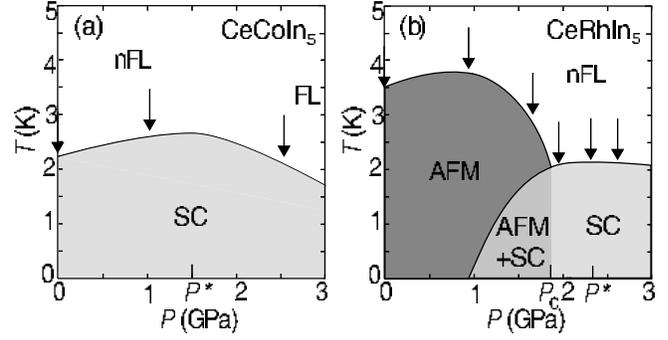}
\caption{Schematic {\it P$-$T} phase diagram for (a) CeCoIn$_5$ and (b) CeRhIn$_5$.    At $P\sim P^{\ast}$, a non-Fermi liquid(nFL)  to Fermi liquid (FL) crossover occurs and  $T_c$ reaches a broad maximum.  In CeRhIn$_5$, $P_c$ is a critical pressure which separates the AF metallic and superconducting states, indicating that the AF QCP is located at or in the vicinity of $P_c$.  In CeCoIn$_5$, the AF QCP appears to be at a  slightly negative inaccessible pressure ($P_c<0$).    Arrows indicate the pressures used in this study.  }
\label{PTPhase}
\end{center}
\end{figure}

Recently, a new class of heavy fermion compounds have been discovered with the chemical fomula Ce$M$In$_5$, where $M$ can be either Co, Rh, or Ir.\cite{pet1,hegger,pet2}.  Ce$M$In$_5$  crystalizes in a quasi two-dimensional (2D) structure that can be viewed as alternating layers of CeIn$_3$ and $M$In$_2$ stacked sequentially along the tetragonal $c$-axis.    At ambient pressure, CeCoIn$_5$ is a superconductor with a transition temperature $T_c$~=~2.3~K.  The electronic specific heat coefficient $\gamma$ have been measured to be $\gamma=$~300~mJ/K$^2$mol at $T_c$.  CeIrIn$_5$ is also a superconductor at ambient pressure with $T_c$~=~0.4~K and $\gamma=$~680~mJ/K$^2$mol.     CeRhIn$_5$ is an AF metal with N\'eel temperature $T_N$~=~3.8~K.  Above $P_c$~=~1.6~GPa, CeRhIn$_5$ undergoes a superconducting transition, indicating that the AF quantum critical point (QCP) is located at or in the vicinity of $P_c$.  Related to the layered structure,  de Haas$-$van Alphen experiments on Ce$M$In$_5$ reveal a corrugated cylindrical Fermi surface \cite{settaisan,shishido}. In addition,  NMR relaxation rate $T_1^{-1}$ measurements indicate the presence of anisotropic (quasi 2D) AF spin fluctuations in the normal state \cite{NMR,NMR2}.

Figures \ref{PTPhase} (a) and (b) display the pressure-temperature ({\it P$-$T} ) phase diagram for CeCoIn$_5$ and CeRhIn$_5$, respectively \cite{sid,kneb,fisher}.  The fgures show that $T_c$ reaches a broad maximum at  characteristic pressures $P^{\ast}\sim 1.5$~GPa for CeCoIn$_5$ and  $P^{\ast}\sim 2.3$~GPa for CeRhIn$_{5}$.   Note that it is reasonable to speculate that the AF QCP of CeCoIn$_5$ is at a slightly negative pressure.  In the region $P<P^{\ast}$,   non-Fermi liquid  behavior is observed in many of the thermodynamic quantities, such as heat capacity\cite{bian}, magnetic susceptibility,\cite{tayama} and NMR relaxation rate $T_1^{-1}$ \cite{NMR,NMR2,NMR3,NMR4}. On the other hand, at $P>P^{\ast}$ the non-Fermi liquid behavior is strongly suppressed and recovery of Fermi liquid behavior is observed. The $P-T$ phase diagram is reminiscent of the phase diagram for high-$T_c$ cuprates around  optimal doping.

Ce$M$In$_5$ superconductors have attracted great interest in the study of strongly correlated electron systems, since the superconductivity instability in Ce$M$In$_5$ occurs in the normal state that exhibits pronounced non-Fermi liquid behavior arising from the proximity of the AF QCP \cite{sid,hegger}.    Among Ce$M$In$_5$ systems, the physical properties of CeCoIn$_5$ have been investigated most extensively.   At ambient pressure, CeCoIn$_5$ shows the hallmarks of proximity to a magnetic instability in the normal state: the logarithmic divergence of the electronic specific heat coefficient $\gamma\sim \ln T$\cite{pet1,ikeda,bian},  Curie$-$Weiss-like temperature dependence of the uniform susceptibility $\chi\sim \frac{1}{T+\theta}$,~\cite{tayama} and a peculiar $T$-dependence of $T_1^{-1}\propto T^{\frac{1}{4}}$~~\cite{NMR,NMR2,NMR3} have been attributed to  strong AF fluctuations in the vicinity of the QCP.    Pronounced non-Fermi liquid behavior is observed in the transport properties, in striking contrast to Eqs.~(1)$-$(3).  Below the coherent temperature $T_{coh}$, which corresponds to the Fermi temperature of $f$-electrons,  down to $T_c$,  $T$-linear dependent $\rho_{xx}$, strongly $T$-dependent $|R_H|$ dramatically increasing with decreasing temperature, and  a striking violation of  Kohler's rule in the magnetoresistance have been  reported.  These unusual properties are markedly pronounced  when the system approaches the AF QCP.   In addition,  the Nernst coefficient $\nu_H$ in the normal state of CeCoIn$_5$ has been reported to be strikingly enhanced by more than three orders of magnitude compared with  conventional metals \cite{bel}.

The superconducting state of Ce$M$In$_5$ has been reported to be highly unusual.  Based on the measurements of specific heat\cite{mov,aoki}, thermal conductivity\cite{izawa}, NMR relaxation rate\cite{NMR3,NMR4}, penetration depth\cite{pene1,pene2}, and Andreev reflection\cite{wei,green},  the superconducting gap function of CeCoIn$_5$ has line nodes and is most likely to have $d_{x^2-y^2}$-wave symmetry\cite{izawa,green,vekhter}, indicating an intimate relation between the superconductivity and the AF fluctuations. Thus the phase diagram of Ce$M$In$_5$, along with the appearence of the $d_{x^2-y^2}$-wave superconductivy near the AF ordered state, indicates that Ce$M$In$_5$ shares some unconventional properties with high-$T_c$ cuprates. Further,  a new pairing state (Fulde-Ferrel-Larkin-Ovchinnikov state) has been suggested at the low-$T$/high-$H$ corner in the $H$-$T$ phase diagram of CeCoIn$_5$.\cite{FFLO}

We stress that the Ce$M$In$_5$ system is very suitable for the detailed study of the charge transport properties of strong correlated electron systems, providing a unique opportunity to clarify the universality of transport properties in the presence of strong AF fluctuations, especially when the system is in the vicinity of the QCP.   This is because {\it the electronic structure can be tuned over a wide range continuously from the AF ordered state to the Fermi liquid state through the non-Fermi liquid state, by applying pressure without introducing additional scattering centers and with keeping the carrier number unchanged.}  In addition,  Ce$M$In$_5$ is in a very clean regime.  In fact, the quasiparticle mean free path in the superconducting state of CeCoIn$_5$ is as long as a few $\mu$m \cite{kasahara}.  This is important because the intrinsic transport properties relevant to the electron$-$electron correlation are often masked bythe impurity scattering.

The case for in Ce$M$In$_5$ contrastz with the chemical doping in high-$T_{c}$ cuprates, which always introduces  randomness that strongly influences the transport properties.  Moreover, in most heavy fermion superconductors, the anomalous Hall term due to  skew scattering, which arises from the asymmetric scattering of conduction electrons by the angular momentum of $f$-electrons, dominates the Hall effect \cite{fert,coleman2,konyam}.  This effect is strongly temperature dependent and  masks the ordinary Hall term arising from the Lorenz force.   In contrast to other heavy fermion compounds, the skew scattering term in Ce$M$In$_5$ is absent or extremely small\cite{nakajima1,nakajima2}.  This allows a detailed analysis of the normal Hall effect to be made.

As well as the non-Fermi liquid behavior of the transport quantities,  the evolution of the Fermi surface when crossing the QCP is an important issue in strongly correlated electron systems.    Two contrary views have been proposed concerning this evolution.    Theories based on spin-density-wave (SDW) predicts a continuous change of the Fermi surface \cite{hertz, millis2}.    On the other hand,  theories based on the breakdown of composite Fermions predict a jump of  the Fermi surface volume  at the QCP\cite{coleman}.   To help clarify this controversial issue,  the importance of  Hall measurements, which can probe the Fermi surface, has been highlighted\cite{coleman}.  Thus CeRhIn$_5$, which can be tune from AF metal to superocnductor,  is a suitable system to study the evolution of the Fermi surface. 

In this paper, we present systematic studies of the dc-resistivity,   Hall effect, and  magnetoresistance of Ce$M$In$_5$ ($M$: Co and Rh),  ranging from the AF metal state to the Fermi liquid state through the non-Fermi liquid state, using pressure as a tuning parameter.   We also show that the transport properties of La$M$In$_5$, which has a similar band structure but has no $f$-electron, exhibit typical Fermi liquid behavior as expressed in Eqs.~(1)$-$(3).  In sharp contrast,  the electron transport properties of CeCoIn$_5$ and CeRhIn$_5$ exhibit a striking deviation from  Fermi liquid behavior.   We analyze these results in accord with the proposed transport theories for non-Fermi liquid behavior in strongly correlated electron systems and compare the results with those for high-$T_c$ cuprates.    Several noticeable features commonly observed in high-$T_c$ cuprates and CeCoIn$_5$ highlight the transport property anomalies in strongly correlated electron systems.  \\

\section{Experimental}

High quality single crystals of CeRhIn$_{5}$, CeCoIn$_{5}$, LaRhIn$_{5}$, and LaCoIn$_{5}$ were grown by the self-flux method.  We performed all measurements on samples with typical dimensions of 1.0$\times$2.0$\times$0.05 mm$^{3}$ in the transverse geometry ( {\boldmath $H$} $\parallel c$, {\boldmath $J$} $\parallel a$).  The Hall effect and transverse magnetoresistance were measured simultaneously.  We obtained $\rho_{xy}$ from the transverse resistance by subtracting the positive and negative magnetic field data.  Hydrostatic pressure up to 2.62 GPa was generated in a piston$-$cylinder type high pressure cell with oil as a transmitting fluid (Daphne 7373). The pressure inside the cell was determined by the superconducting transition temperature of Pb.  \\

\section{Results}

\subsection{dc-resistivity}

\begin{figure}[t]
\begin{center}
\includegraphics[width=9cm]{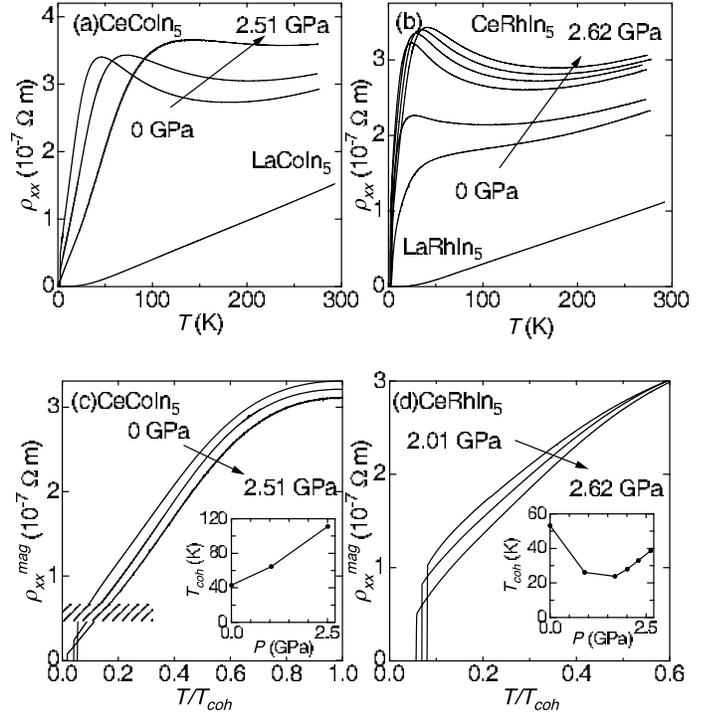}
\caption{(a) Temperature dependence of the dc-resistivity $\rho_{xx}$ in zero field of CeCoIn$_5$ under different pressures, together with the data of LaCoIn$_5$.  (b) The same plot for CeRhIn$_5$ and LaRhIn$_5$. (c) Magnetic part of the resistivity, $\rho^{mag}_{xx}=\rho_{xx}-\rho_{xx}({\rm LaCoIn_5})$, for CeCoIn$_5$ obtained from (a), plotted as a function of $T/T_{coh}$.  The hatched area indicates the region where $R_H$ is minimum.  For detail, see the text (\S4.4).  Inset:  $P$-dependence of $T_{coh}$, at which $\rho_{xx}^{mag}$ is maximum, for CeCoIn$_5$.  (d) $\rho^{mag}_{xx}=\rho_{xx}-\rho_{xx}({\rm LaRhIn_5})$ for CeRhIn$_5$ obtained from (b), plotted as a function of $T/T_{coh}$.   Inset: $P$-dependence of $T_{coh}$ for CeRhIn$_5$.  }
\label{DCres}
\end{center}
\end{figure}

Figures \ref{DCres} (a) and (b) depict the temperature dependence of the dc-resisticity $\rho_{xx}$ in zero field under pressure for CeCoIn$_5$ and CeRhIn$_5$, respectively.  $\rho_{xx}$ of  LaCoIn$_5$ and LaRhIn$_5$ are also plotted.   The overall behavior of Ce$M$In$_5$ is typical for  Ce-based heavy fermion compounds.  In the high temperature regime where the $f$-electrons are well localized,  $\rho_{xx}$ increases gradually with decreasing temperature due to the Kondo effect between the conduction electrons and localized $f$-electron spins.   Associated with the coherent$-$incoherent crossover, $\rho_{xx}$ shows a broad maximum around the coherent temperature $T_{coh}$, and at lower temperatures where the $f$-electrons become itinerant, $\rho_{xx}$ exhibits a metallic temperature dependence.

Figures \ref{DCres}~(c) and (d) depict the magnetic part of the resistivities obtained by subtracting resistivity of La$M$In$_5$ from $\rho_{xx}$,   $\rho_{xx}^{mag} \equiv \rho_{xx}-\rho_{xx}({\rm La}M{\rm In_5})$, plotted as a function of $T/T_{coh}$ in the superconducting regime.  Here $T_{coh}$ is defined as the temperature at which $\rho_{xx}^{mag}$ is maximum.   The insets of Figs.~\ref{DCres}~(c) and (d) show the pressure dependence of $T_{coh}$.    An increase in pressure leads to an increase in $T_{coh}$  and a decrease in $\rho_{xx}$ due to an enhancement of the hybridization between conduction electrons and $f$-electrons through the expansion of the band width of the conduction electrons.  On the other hand, $T_{coh}$ decreases with pressure in the AF metal regime of CeRhIn$_5$ ($P<P_c$ in Fig.~1~(b)).

At ambient pressure, $\rho_{xx}^{mag}$ of CeCoIn$_5$ displays an almost perfect linear $T$-dependence below $T \sim T_{coh}/2$ down to $T_c$.  At $P$ = 2.51~GPa, however, an apparant deviation from $T$-linear dependence is seen in $\rho_{xx}^{mag}$; the overall $T$-dependence above $T_c$ up to $T\sim T_{coh}/2$  is well fitted by $\rho_{xx}^{mag} \propto T^{\alpha}$ with $\alpha \approx$ 1.2.  For CeRhIn$_5$, the resistivity shows $T$-dependence with a convex shape with $\alpha <1$ above $T_c$.  This behavior is less pronounced as the pressure increases.  The fact that the power $\alpha$ in the temperature dependence of the resistivity increases  with pressure is an indication that the system approaches  the Fermi liquid regime. \\

\subsection{Hall effect}

\begin{figure}[t]
\begin{center}
\includegraphics[width=7.0cm]{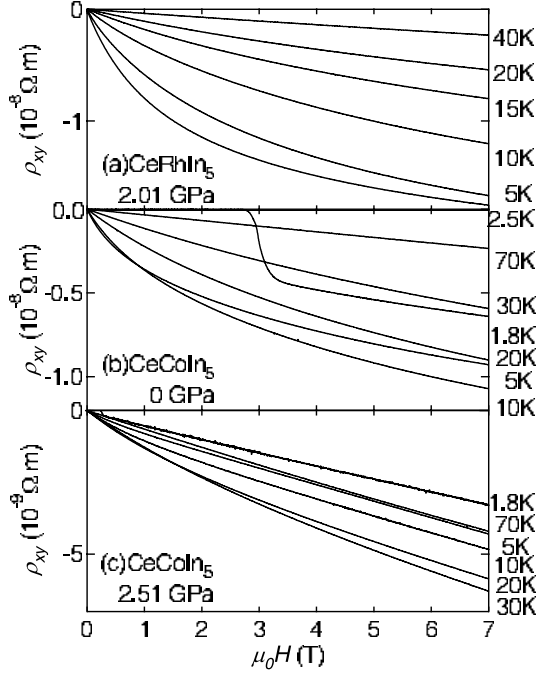}
\caption{Field dependence of the Hall resistivity $\rho_{xy}$ for (a) CeRhIn$_5$ at $P$~=~2.01~GPa ($P>P_c$), (b) CeCoIn$_5$ at ambient pressure, and (c) CeCoIn$_5$ at $P$~=~2.51~GPa.  }
\label{Hall1}
\end{center}
\end{figure}

Figure \ref{Hall1} (a) depicts the $H$-dependence of the Hall resistivity $\rho_{xy}$ for CeRhIn$_5$ at $P$~=~2.01~GPa  ($P>P_c$).  Figures \ref{Hall1} (b) and (c) depict $\rho_{xy}(H)$  for CeCoIn$_5$ at ambient pressure and at $P$~=~2.51~GPa (see Figs.~\ref{PTPhase} (a) and (b)).  The Hall sign is negative for both systems in the present $T$-, $H$-, and $P$-ranges.  At low temperatures, $\rho_{xy}$ exhibits a non-linear $H$-dependence, which is more pronounced when the system approaches the AF QCP.  

\begin{figure}[t]
\begin{center}
\includegraphics[width=7.5cm]{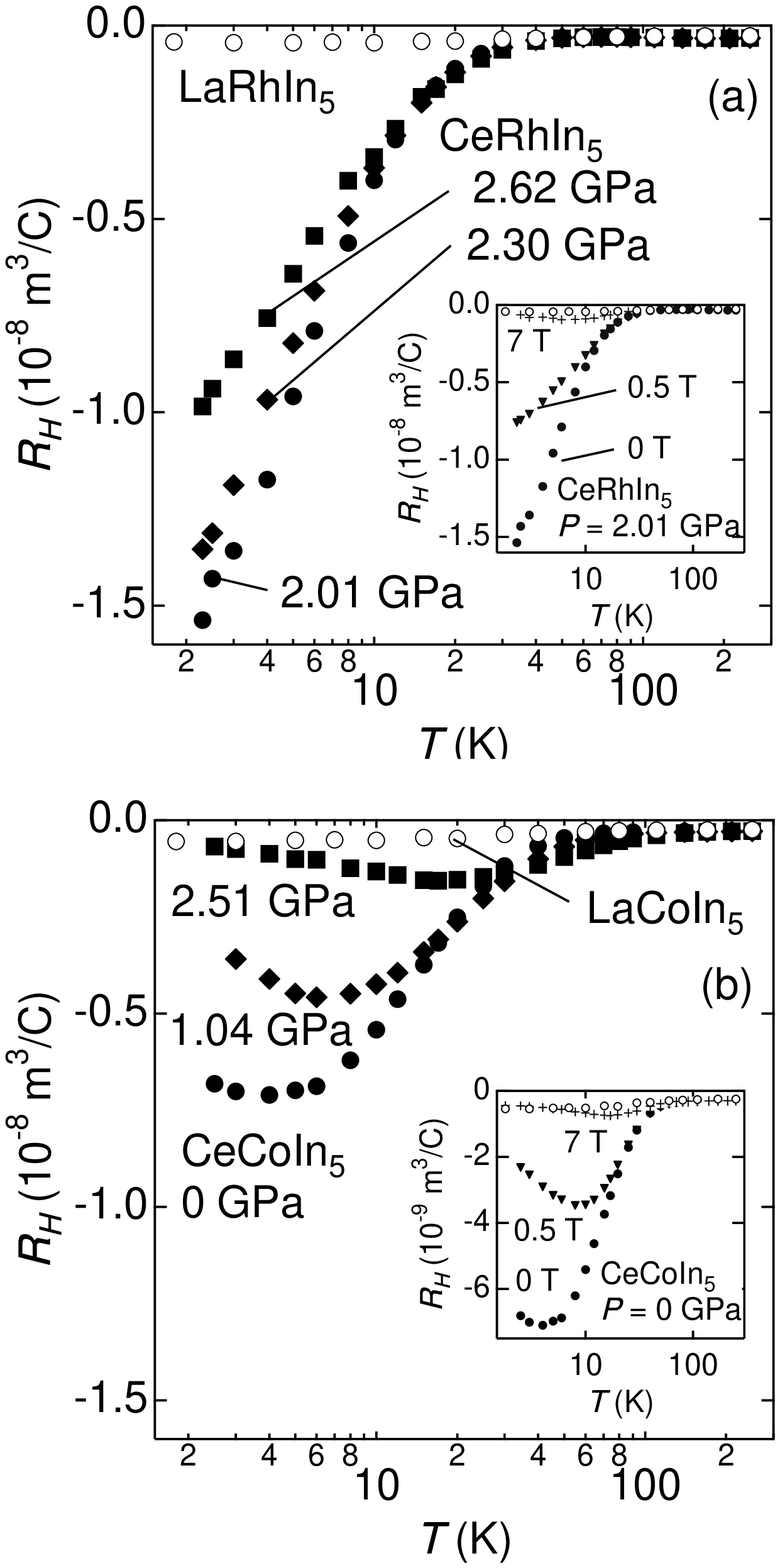}
\caption{(a) Temperature dependence of $R_H$ for CeRhIn$_5$ in the superconducting regime at $P>P_c$,  $P$ = 2.01 ($\bullet$), 2.30 ($\blacklozenge$),  and 2.62~GPa ($\blacksquare$) at  the zero field limit, $\lim_{H\rightarrow0}\frac{{\rm d}\rho_{xy}}{{\rm d}H}$, and for LaRhIn$_5$ at $P$~=~0 ($\circ$).  Inset:  The same data for CeRhIn$_5$ at $P$~=~2.01~GPa at $\mu_0H=$~0~T ($\bullet$), 0.5~T ($\blacktriangledown$),  and  7~T ($+$),  defined as ${\rm d}\rho_{xy}/{\rm d}H$,  and  for LaRhIn$_5$ ($\circ$).   (b) Temperature dependence of $R_H$ for CeCoIn$_5$ at $P$ = 0 ($\bullet$), 1.04 ($\blacklozenge$),  and 2.51~GPa ($\blacksquare$) at  the zero field limit, $\lim_{H\rightarrow0}\frac{{\rm d}\rho_{xy}}{{\rm d}H}$, and for LaCoIn$_5$ ($\circ$).  Inset:  The same data for CeCoIn$_5$ at ambient pressure at $\mu_0H=$~0~T ($\bullet$), 0.5~T ($\blacktriangledown$),  and  7~T ($+$),  defined as ${\rm d}\rho_{xy}/{\rm d}H$,  and  for LaCoIn$_5$ ($\circ$). 
  }
\label{Hall2}
\end{center}
\end{figure}

We first discuss the Hall effect at $P>P_c$ where the system undergoes a superconducting transition.   Figures \ref{Hall2} (a) and (b) depict the $T$-dependence of $R_H$, defined as the field derivative of $\rho_{xy}$,  $R_H\equiv\frac{{\rm d}\rho_{xy}}{{\rm d}H}$,  at the zero field limit ($H \rightarrow 0$) for CeRhIn$_5$ at $P>P_c$ and for CeCoIn$_5$, respectively.   For comparison, $R_H$ for LaRhIn$_5$ and LaCoIn$_5$ at ambient pressure is also plotted in Figs. \ref{Hall2} (a) and (b), respectively.    Although we do not show it here, $\rho_{xy}$ for both La-compounds show nearly perfect $H$-linear dependence.   With decreasing $T$, $R_H$ for both La-compounds show a slight increase and the amplitude at low temperature is approximately $-5 \times 10^{-10}$ m$^3$/C, which is nearly half the Hall coefficient expected for one electron per unit cell.  This indicates the presence of small hole pockets at the Fermi surface.   

Before discussing the Hall effect in Ce$M$In$_5$, we recall the Hall effect in other heavy fermion systems.   The Hall effect in conventional heavy fermion compounds has been studied extensively and is known to exhibit a similar behavior, irrespective of the system  (see Figs.~1 and 2 in Ref. [\citen{fert}]).   At high temperatures above $T_{coh}$, $R_H$ is mostly positive and is much larger than $1/ne$.   Below $T_{coh}$, $R_H$  decreases rapidly with decreasing $T$ after showing a broad maximum at a temperature slightly below $T_{coh}$.  At very low temperature, the amplitude of $R_H$ is strongly reduced and becomes nearly $T$-independent.  This anomalous temperature dependence of $R_H$ can be explained as follows.   Generally the Hall effect in heavy fermion system can be decomposed into two terms: $R_H=R_H^n+R_H^a$.  Here, $R_H^n$ is the ordinary Hall effect due to the Lorentz force and is nearly $T$-independent within the Bloch$-$Boltzmann approximation.  Meanwhile, $R_H^a$ represents the so-called ``anomalous Hall effect''  due to skew scattering, which  originates from the assymetric scattering of the conduction electrons by the angular momenta of $f$-electrons, induced by the external magnetic field.  The sign of $R_H^a$ is positive in almost all Ce-based heavy fermion compounds.  This term is strongly temperature-dependent and is well-scaled by the uniform susceptibility: above $T_{coh}$, $R_H^a$ scales as $R_H^a \propto \chi$ \cite{konyam} or $\propto \chi \rho_{xx}$ \cite{fert,coleman2}, while below $T_{coh}$, $R_H^a \propto \chi \rho_{xx}2$ \cite{konyam}.
The magnitude of $R_H^a$ is much larger than $|R_H^n|$ except at  $T \ll T_{ coh}$ and $T \gg T_{ coh}$, where the contribution of the skew scattering vanishes.

\begin{figure}[t]
\begin{center}
\includegraphics[width=7.5cm]{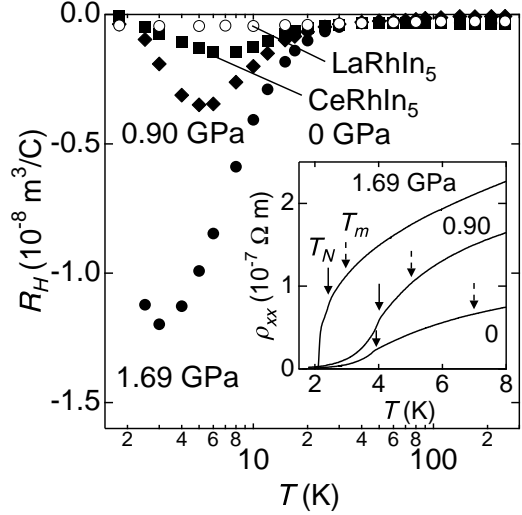}
\caption{Main panel: Temperature dependence of $R_H$ for CeRhIn$_5$ in the AF metallic regime at $P<P_c$,  $P$ = 1.69 ($\bullet$), 0.90 ($\blacklozenge$),  and 0~GPa ($\blacksquare$) at the zero field limit, and for LaRhIn$_5$ at $P$~=~0 ($\circ$).  Inset:  Temperature dependence of the dc-resistivity for CeRhIn$_5$ in the AF metallic regime.  Solid arrows indicate the N\'eel temperature. Dashed arrows indicate the temperature at which the amplitude of $R_H$ is maximum. }
\label{Hall3}
\end{center}
\end{figure}

\begin{figure}[b]
\begin{center}
\includegraphics[width=7.5cm]{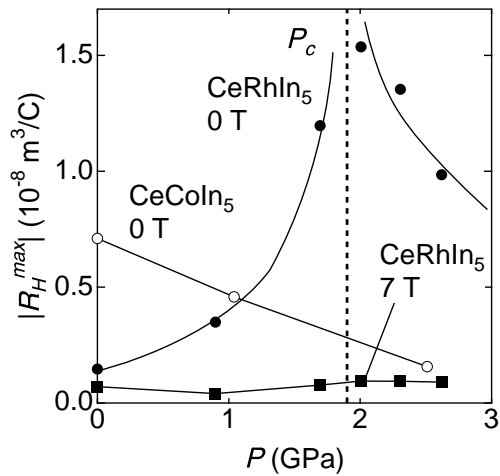}
\caption{Maximum of $|R_H|$ vs. pressure for CeRhIn$_5$ at $\mu_0H=$~0~T ($\bullet$), and  7~T ($\blacksquare$) and for CeCoIn$_5$ at $\mu_0H=$~0~T ($\circ$).  At $P_c$, $|R_H|$ of CeRhIn$_5$ at 0~T exhibits a divergent behavior. }
\label{Hall4}
\end{center}
\end{figure}

It is obvious that the Hall effect in CeRhIn$_5$ and CeCoIn$_5$ shown in Figs.~\ref{Hall2} (a) and (b) is distinctly different from those in other heavy fermion systems.  The first important signature is that $R_H$ is nearly $T$-independent in the high temperature regime above $T_{coh}$ and cannot be scaled either by $\chi\rho_{xx}$ or  $\chi$, both of which have a strong $T$-dependence above $T_{coh}$.   In addition,  $\chi$ is almost strictly $H$-linear up to 12~T at $T>$~2~K, while $\rho_{xy}$ shows a nonlinear $H$-dependence,  as shown in Figs.~\ref{Hall1} (a), (b), and (c).   Moreover, the amplitude of  $R_H$ above $T_{coh}$ is much smaller than those of other heavy fermion systems above $T_{coh}$.   Thus, the sign,  amplitude, and  $T$-dependence of $R_H$ of CeRhIn$_5$ and CeCoIn$_5$ are clearly in conflict with the conventional skew scattering mechanism.    These results prove that  skew scattering in CeRhIn$_5$ and CeCoIn$_5$ is negligibly small and that $R_H$ is dominated by the normal part of the Hall effect.    At present, it is not clear why skew scattering is absent in Ce$M$In$_5$ is not clear and a theoretical study based on  realistic $p$-$f$ model for CeMIn$_5$ would be required to clarify this point.  The absence of skew scattering enables us to analyze the normal Hall effect $R_H^n$ in detail.

In both CeRhIn$_5$ and CeCoIn$_5$, the temperature dependence of the Hall effect is closely correlated with that of the resistivity.   At high temperatures $T>T_{coh}$,  $R_{H}$ at all pressures in Ce$M$In$_5$ is nearly independent of temperature and coincides well with $R_H$ of La$M$In$_5$, {\it i.e.} $R_H\sim1/ne$.  This is consistent with band structure calculations, which predict that the electronic structure of Ce$M$In$_5$ is similar to that of La$M$In$_5$.    Therefore it is natural to assume that  $R_H$ of La$M$In$_5$ is close to the Hall coefficient in the Fermi liquid limit of Ce$M$In$_5$.  Below $T_{coh}$,  the carrier number of Ce$M$In$_5$ is expected to  slightly increase from that of La$M$In$_5$,  since the $f$-electrons participate in the conduction.    

Below $T_{coh}$,  $R_H$ of both CeRhIn$_5$ and CeCoIn$_5$ exhibits a remarkable departure from $R_H$ of LaRhIn$_5$ and LaCoIn$_5$, respectively; the amplitude of $R_H$ increases rapidly below $T_{coh}$ at all pressures  as the temperature is lowered.   At very low temperatures, the behavior of $R_H$ is different for both CeRhIn$_5$ and CeCoIn$_5$.  In CeRhIn$_5$, as shown in Fig.~\ref{Hall2}~(a), the ampltude of $R_H$ increases monotonically down to $T_c$.  Below $T_c$, the Hall signal vanishes.   At $P$~=~2.01~GPa, slightly above $P_c$, the amplitude of $R_H$  is strikingly enhanced below $T_{coh}$;  the amplitude of $R_H$ just above $T_c$ is enhanced to nearly 50 times that at high temperatures above $T_{coh}$, which is $\sim|1/ne|$.   On the other hand,  in CeCoIn$_5$, $R_H$ exhibits an upturn at low temperatures; the aplitude of $R_H$ decreases after showing a maximum at around $T$~=~4, 6, and 20~K for $P$ = 0, 1.04, and 2.51~GPa, respectively.   At ambient pressure, the amplitude of $R_H$ at low temperatures is enhanced to nearly 30 times that above $T_{coh}$.  In both CeRhIn$_5$ and CeCoIn$_5$, the low temperature enhancement of the amplitude of $R_H$ is dramatically suppressed in the high presure regime where non-Fermi liquid behavior is  suppressed.  Interestingly,  $R_H$ of CeCoIn$_5$ at $P$ = 2.51~GPa approaches  $R_H$ of LaCoIn$_5$, {\it i.e.} the Fermi liquid value,  at very  low temperatures.

\begin{figure}[t]
\begin{center}
\includegraphics[width=8cm]{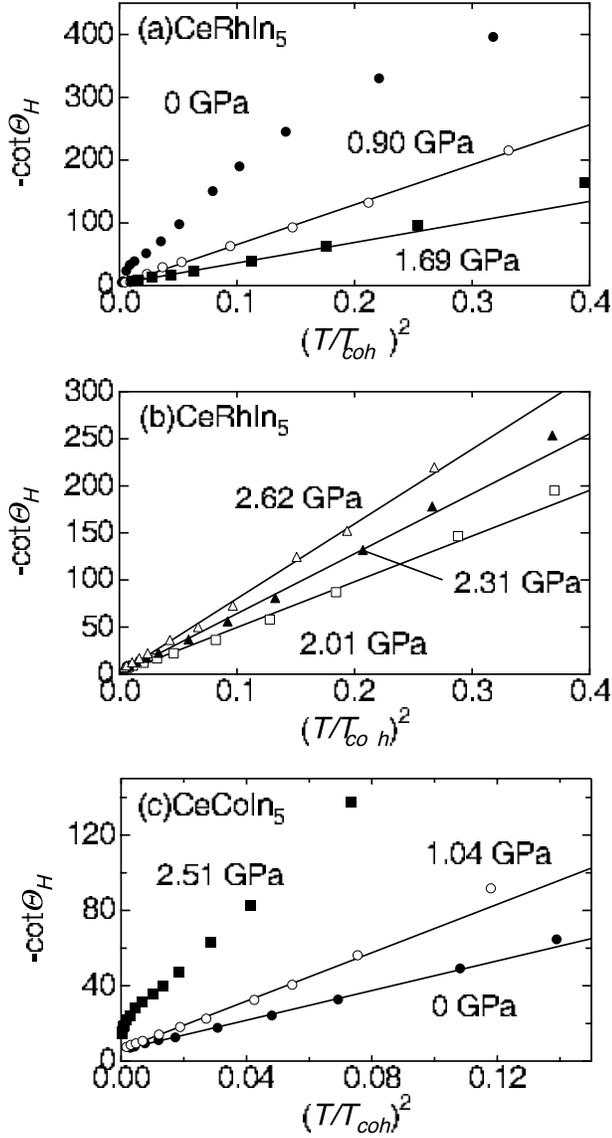}
\caption{$-\cot \Theta_H$ versus $T/T_{coh}$ for (a) CeRhIn$_5$ in the AF metallic regime ($P<P_c$), (b) CeRhIn$_5$ in the superconducting regime ($P>P_c$), and (c) CeCoIn$_5$, at different pressures.  The solid lines represent the $T^2$-dependence of $\cot \Theta_H$. }
\label{Hall5}
\end{center}
\end{figure}

We next discuss the Hall effect of CeRhIn$_5$ in the AF metallic regime at $P<P_c$ (see Fig.~1~(b)).   Figure~\ref{Hall3} shows the $T$-dependence of $R_H$.    Below $T_{coh}$, as the temperature is lowered,  the amplitude of $R_H$ increases rapidly and then decreases after a maximum at $\sim$3~K, $\sim$5~K, and $\sim$7~K for $P$~=~1.69, 0.90 and 0~GPa, respectively. This behavior is similar to that for CeCoIn$_5$ shown in Fig.~\ref{Hall2}~(b).   The inset of  Fig.~\ref{Hall3} depicts the $T$-dependence of $\rho_{xx}$ at low temperatures.    Below the N\'eel temperature $T_N$, shown by solid arrows,  $\rho_{xx}$ varies as $T^2$, indicating  Fermi liquid behavior.   The dashed arrows indicate the temperature at which the amplitude of $R_H$ is maximum.  These temperatures are located well above $T_N$.

Thus there are three distinct $T$-regions in the Hall effect: high-$T$-region, where $R_H$ is nearly $T$-independent; intermediate-$T$-region, where the amplitude of $R_H$ increases rapidly ; and low-$T$-region, where the amplitude of $R_H$ begins to decrease.  (A schematic figure of  $R_H$  is illustrated in Fig.~\ref{HallT}.  We denote the three regions as region-(I),  (II), and (III).)  The low-$T$-region is absent in CeRhIn$_5$ in the superconducting region at $P>P_c$.  We will discuss the behavior of the Hall effect in the low-$T$-region in a later section.

In Fig.~\ref{Hall4}, we plot the $P$-dependence of the maximum values of $|R_H|$, $|R_H^{\max}|$,  obtained from Figs.~\ref{Hall2}~(a) and (b) and Fig.~\ref{Hall3}.     For CeRhIn$_5$,  $|R_H^{\max}|$ at $\mu_0H\rightarrow$ 0~T increases rapidly as $P_c$ is approached from either the AF metallic ($P<P_c$) or superconducting ($P>P_c$) regimes.  At $\mu_0H$~=~7T, $|R_H^{\max}|$ exhibits a maximum at a pressure slightly higher than $P_c$.  We will discuss this behavior later.   For CeCoIn$_5$ with negative $P_c$, $|R_H^{\max}|$ increases as $P_c$ is approached.    {\it These results provide  direct evidence that the Hall effect is strongly influenced by the AF fluctuations associated with the AF QCP.}

It has been argued that in high-$T_c$ cuprates,  the Hall problem can be simplified when analyzed in terms of the cotangent of the Hall angle $\cot \Theta_H$ \cite{ando,chien,anderson,hussey}.  A very recent systematic study of $R_H$ on La$_{2-x}$Sr$_x$CuO$_4$ single crystals over a wide doping range has shown that the $T^2$-law of $\cot \Theta_H$ holds well from the underdoped region up to optimal doping, but gradually breaks down,   showing a negative curvature when the system is overdoped \cite{ando}.   We here examine the pressure dependence of  $\cot \Theta_H$ for CeRhIn$_5$ and CeCoIn$_5$.  Figures~\ref{Hall5} (a) and (b) depict $\cot \Theta_H$ for CeRhIn$_5$ at $P<P_c$ and $P>P_c$, respectively.  We plot the data as a function of  $(T/T_{coh})^2$.  Figure~\ref{Hall5} (c) shows the same plot for CeCoIn$_5$.    As seen from the solid lines in these figures,   $\cot \Theta_H$ obeys a $T^2$-dependence near the AF QCP.   On the other hand, $\cot \Theta_H$ apparently deviates from the $T^2$-law with a negative curvature,  as shown for  CeCoIn$_5$ at $P$~=~2.51~GPa in the regime where the non-Fermi liquid behavior is less pronounced.  This behavior is similar to overdoped La$_{2-x}$Sr$_x$CuO$_4$.  For CeRhIn$_5$ at ambient pressure, deviation from the $T^2$-law is also observed.

We here comment on the recent consideration of $R_H$ in CeCoIn$_5$  in terms of the two-fluid Kondo model,  in which  the Kondo lattice system is divided into coherent and incoherent parts.   It has been proposed that $R_H$ can be expressed as $R_H=\alpha_f R_H^{La}$, where $\alpha_f$ is the $f$-electron weighting function reflecting the coherent part  \cite{lahall}.  However, if we apply this model to $R_H$ at 2.51~GPa,  $\alpha_f$ is close to unity at low temperatures, indicating that there is no coherent part.   This is not consistent with itinerant $f$-electrons in this temperature regime.   Therefore the two-fluid Kondo scenario appears to be inconsistent with the present results.\\

\subsection{Magnetoresistance}

\begin{figure}[b]
\begin{center}
\includegraphics[width=7.5cm]{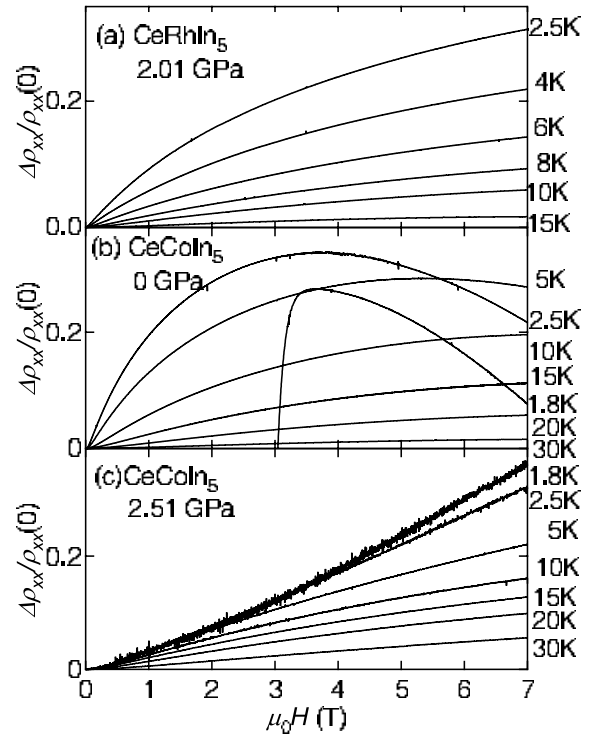}
\caption{ Magnetoresistance $\Delta\rho_{xx}/\rho_{xx}$ at different temperatures for (a) CeRhIn$_5$ at $P$~=~2.01~GPa ($P>P_c$), (b) CeCoIn$_5$ at ambient pressure, and (c) CeCoIn$_5$ at $P$~=~2.51~GPa.}
\label{MR1}
\end{center}
\end{figure}

Figure \ref{MR1} (a) displays the magnetoresistance $\Delta \rho_{xx}/\rho_{xx}$ at various temperatures for CeRhIn$_5$ at $P$~=~2.01~GPa ($P>P_c$).    The same plot for CeCoIn$_5$ at $P$~=~0 and 2.51~GPa is shown in Figs. \ref{MR1} (b) and (c), respectively.   The magnetorsistance is positive at low fields for both compounds.   The  magnetoresistance exhibits $H^2$-dependence at very low fields; for CeCoIn$_5$ at ambient pressure, this field range is less than 0.1~T.    For CeRhIn$_5$ at 2.01~GPa and CeCoIn$_5$ at ambient pressure, where the system is close to the AF QCP, the resistivity shows $H$-dependence with a concave downward shape at low temperatures.    At ambient pressure for CeCoIn$_5$, the slope of the magnetoresistance turns from positive to negative at high fields  below 5 K.   This crossover field increases with temperature.  The magnitude of the low field positive magnetoresistance decrease with temperature.  The origin of high field negative magnetoresistance is possibly due to suppression of spin-flop scattering by the magnetic field, which is often observed in conventional heavy fermion compounds below $T_{coh}$.   In this paper, we will address the low field magnetoresistance.

\begin{figure}[b]
\begin{center}
\includegraphics[width=8cm]{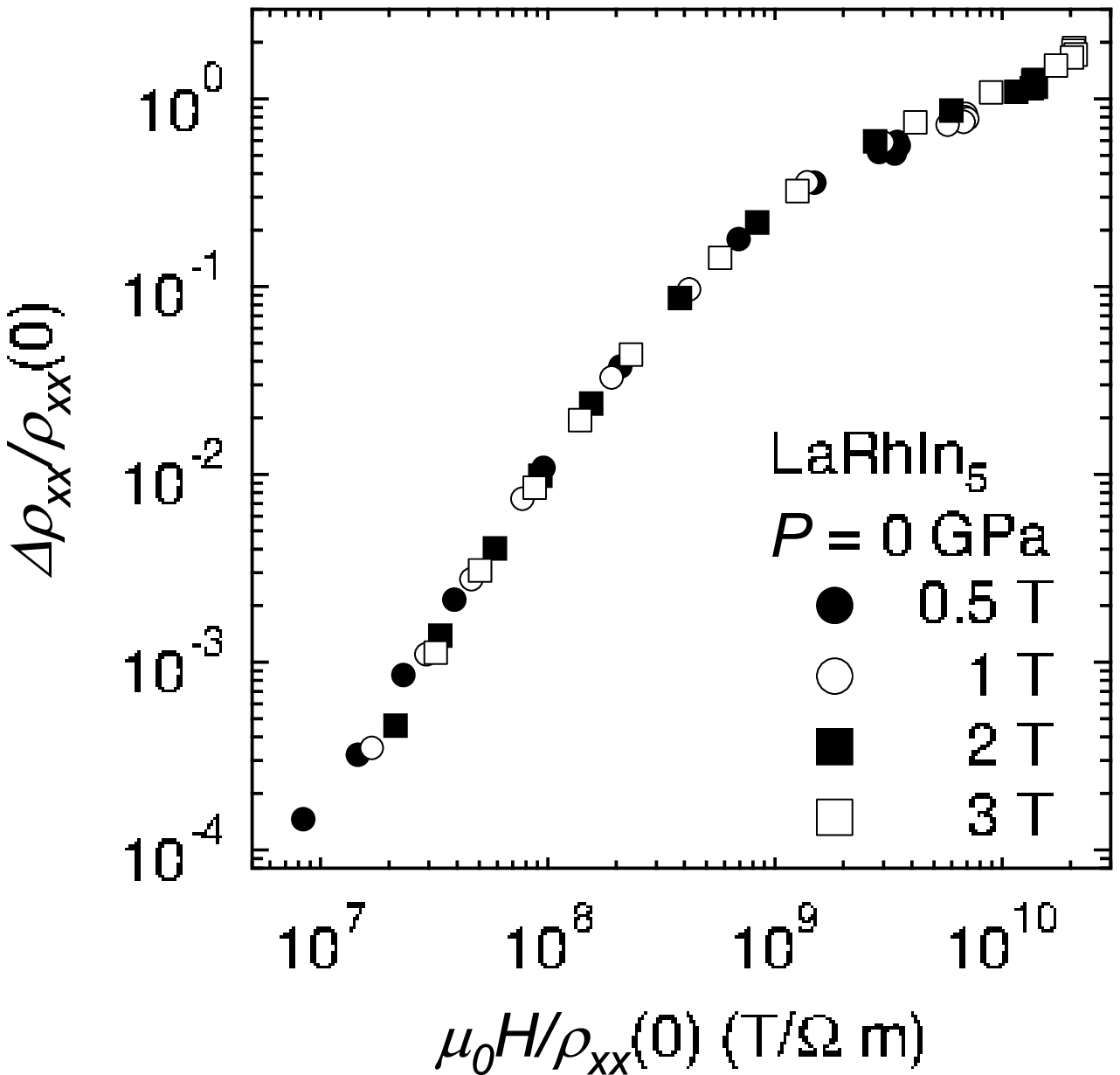}
\caption{Kohler's plot: $\Delta \rho_{xx} / \rho_{xx}$ versus $\mu_0H/\rho_{xx}$ for LaRhIn$_5$ at ambient pressure.    $\Delta\rho_{xx} / \rho_{xx}$ at different fields collapses into the same curve for four orders of magnitude, indicating the applicability of the Kohler's rule shown in Eq.~(3)}
\label{MR2}
\end{center}
\end{figure}

\begin{figure*}[t]
\begin{center}
\includegraphics[width=15cm]{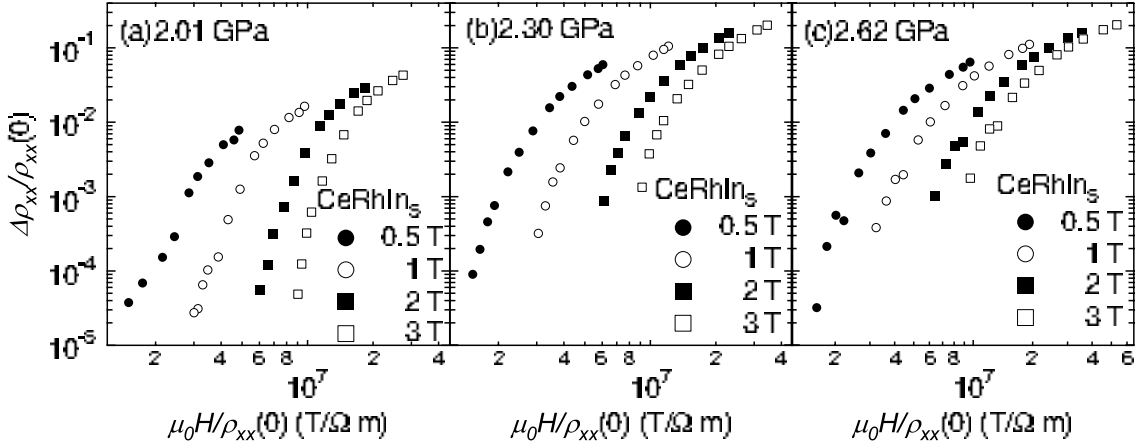}
\caption{Kohler's plot: $\Delta\rho_{xx}/\rho_{xx}$ versus $\mu_0H/\rho_{xx}$ for CeRhIn$_5$ in the superconducting regime ($P>P_c$) at (a) $P$~=~2.01, (b) 2.30, and (c) 2.62~GPa.   Apparent violation of  Kohler's rule is observed at all pressures.    }
\label{MR3}
\end{center}
\end{figure*}

\begin{figure*}[t]
\begin{center}
\includegraphics[width=15cm]{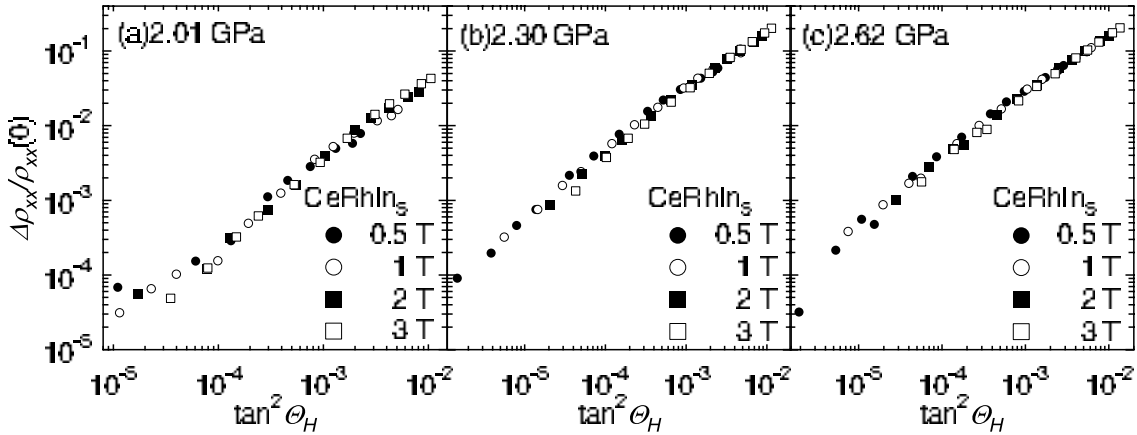}
\caption{ $\Delta\rho_{xx}/\rho_{xx}$ plotted as a function of  $\tan^2 \Theta_H$ for CeRhIn$_5$ in the superconducting regime ($P>P_c$) at (a) $P$~=~2.01, (b) 2.30, and (c) 2.62~GPa.   At all pressures,  the magnetoresistance at different fields collapses into the same curves for four orders of magnitude in $\Delta \rho_{xx}/\rho_{xx}(0)$, indicating the applicability of the modified Kohler's rule shown in Eq.~(6). }
\label{MR4}
\end{center}
\end{figure*}

In conventional metals, a positive magnetoresistance usually obeys Kohler's rule, a scaling relation described by Eq.~(3).  Violation of Kohler's rule is sometimes observed even in conventional metals when the Fermi surface is composed of multiple bands and  the carrier scattering rate is strongly band dependent\cite{pippard}.    To examine Kohler's rule in the present systems,  it is necessary to extract the orbital part of the magnetoresistance.  Usually, for this purpose, the longitudinal magnetoresistance measured in the geometry {\boldmath $H$} $\parallel $ {\boldmath $J$} $\parallel a$ is subtracted from the transverse magnetoresistance.  However, in a material with a 3D Fermi surface, the orbital magnetoresistance appears even in the longitudinal geometry \cite{ziman}.   We therefore assume that the transverse magnetoresistance is dominated by the orbital components in the present materials.    To check this, we test Kohler's rule for LaRhIn$_5$, which has a similar band structure as CeRhIn$_5$.    Figure~\ref{MR2} displays Kohler's plot for LaRhIn$_5$; $\Delta \rho_{xx}/\rho_{xx}(0)$  plotted as a function of $\mu_0H/\rho_{xx}(0)$.     $\Delta \rho_{xx}/\rho_{xx}(0)$ at different fields collapses into the same curve for four orders of magnitude as a function of $\mu_0H/\rho_{xx}(0)$, indicating that the magnetoresistance obeys the Kohler's rule in Eq.~(3) quite well.   We obtained similar results for LaCoIn$_5$.  These results indicate the validity of the assumptions that the multiband effect is not important and the transverse magnetoresistance is dominated by orbital effects.

Figures~\ref{MR3} (a), (b), and (c) shows Kohler's plots for CeRhIn$_5$ at $P$~=~2.01, 2.30, and 2.62~GPa ($P>P_c$ in Fig.~1~(b)), respectively.  In sharp contrast to LaRhIn$_5$,  $\Delta \rho_{xx}/\rho_{xx}(0)$ at different fields are on  distinctly different curves, indicating apparent vaiolation of Kohler's rule.    We next examine the data in accordance with Eq.~(6),  which has shown to describe well the magnetoresistance in high-$T_c$ cuparates\cite{harris}.  Figures ~\ref{MR4} (a), (b), and (c) display the magnetoresistance as a function of $\tan^2\Theta_H$.  {\it At all pressures,  the magnetoresistance data at different fields collapse into the same curves for four orders of magnitude in $\Delta \rho_{xx}/\rho_{xx}(0)$.}    Although we do not show it here, we obtained essentially the same results for CeCoIn$_5$.  Thus in the pressure regime where non-Fermi liquid behavior is pronounced in CeCoIn$_5$ and CeRhIn$_5$, the magnetorsesistance exhibits a striking violation of Kohler's rule in Eq.~(3),  while the magnetorsistance obeys the scaling relation expressed by Eq.~(6)quite well.

\begin{figure}[t]
\begin{center}
\includegraphics[width=7.5cm]{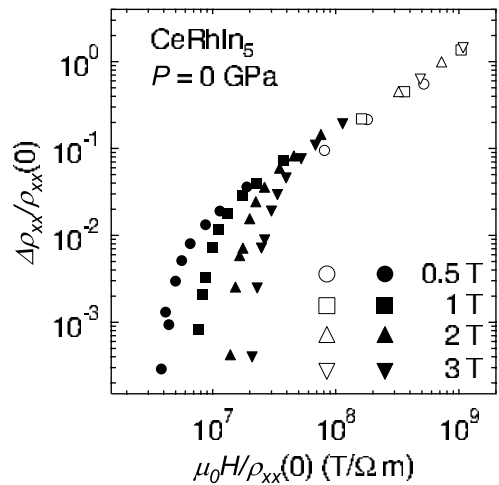}
\caption{Kohler's plot, $\Delta\rho_{xx}/\rho_{xx}$ versus $\mu_0H/\rho_{xx}$ in the AF metal regime ($P<P_c$) of CeRhIn$_5$.  The solid and open circles represent the data above and below $T_N$.    Kohler's rule is satisfied below $T_N$, where the resistivity shows a Fermi liquid behavior, shown in the inset of Fig.~\ref{Hall3}.  In sharp contrast violation of Kohler's rule is observed above $T_N$, where non-Fermi liquid behavior is observed in the resistivity.  }
\label{MR5}
\end{center}
\end{figure}

Interestingly,  both  violation and nonviolation of Kohler's rule can be seen simultaneously in the AF metal regime ($P<P_c$) of CeRhIn$_5$, as shown in Fig.~\ref{MR5}.  The solid and open circles represent the data above and below $T_N$.   As shown in the inset of Fig.~\ref{Hall3}, $\rho_{xx}$ exhibits Fermi liquid behavior below $T_N$, while non-Fermi liquid behavior is observed in $\rho_{xx}$ above $T_N$.   The magnetoresistance well obeys Kohler's rule below $T_N$, while  violation of  Kohler's rule is observed above $T_N$.   These results reinforce the conclusion that  Kohler's rule is violated only in the regime where non-Fermi liquid behavior is observed.   \\

\section{Discussion}

\subsection{Striking resemblance between Ce$M$In$_5$ and high-$T_c$ cuprates}

Summarizing the non-Fermi liquid behavior observed in the transport properties of Ce$M$In$_5$ below the Fermi temperature of $f$-electrons $T_{coh}$, 

\begin{enumerate}
\item The dc-resistivity shows $T$-linear dependence, $\rho_{xx}\propto T$,  over a wide temperature range below $T_{coh}$ down to $T_c$.
\item  Below $T_{coh}$, $|R_H|$  increases rapidly with decreasing temperature, depending on $T$ as $R_H \propto 1/T$,  and reaches a value much larger than $|1/ne|$ at low temperatures.    The  cotangent of the Hall angle  varies as $\cot\Theta_H\propto T^2$.   
\item The magnetoresistance displays $T$- and $H$-dependence that strongly violate Kohler's rule, and  is well scaled by the tangent of the Hall angle: $\Delta\rho_{xx}/\rho_{xx} \propto \tan^2\Theta_H$.
\end{enumerate}
We emphasize that all of these salient features have been reported for high-$T_c$ cuprates, as discussed in \S1.  {\it The striking resemblance between high-$T_c$ cuprates and Ce$M$In$_5$ lead us to consider that the anomalous transport properties in both systems originate from the same mechanism.}  It should be noted that in high-$T_c$ cuprates the presence of the pseudogap in the underdoped regime reduces both $R_H$ and $\rho_{xx}$, while the presence of  the pseudogap at $P<P^{\ast}$ in CeCoIn$_5$ is controversial\cite{sid}.   Since the anomalous transport behavior seem to present an insurmountable challenge to the Fermi liquid theory, several notable attempts to explain these  behavior have been made for high-$T_c$ cuprates.   Two broad  approaches have been proposed to understand  the transport properties in high-$T_c$ cuprates:   non-Fermi liquid and Fermi liquid approaches.

In models based on a non-Fermi liquid ground state,  it is proposed that the transport carriers are exotic objects, such as spinons and holons that may model the doped CuO$_2$ 2D planes \cite{chien,anderson,harris}.  This model involves two distinct relaxation times for momentum changes parallel and perpendicular to the Fermi surface, associated with spin$-$charge separation.    It has been argued that the $T$-linear $\rho_{xx}$ and the  $T^2$-dependence of  $\cot \Theta_H$  imply the existence of different longitudinal and Hall relaxation rates that change as $\tau_{tr} \propto T$ and $\tau_{H}\propto T^2$, respectively.  Then, the diagonal conductivity, Hall conductivity, and magnetoresistance depend on $\tau_H$ and $\tau_{tr}$ as $\sigma_{xx} \propto \tau_{tr}$, $\sigma_{xy} \propto \tau_H \tau_{tr}$, and $\Delta \rho_{xx} / \rho_{xx} \propto \tau_H^{-2}$, respectively, which can reproduce the data.  However this idea of scattering rate separation has not yet gained  complete consensus.   It should be noted that the parent material of high-$T_c$ cuprates is a Mott insulator, while that of Ce$M$In$_5$ is an AF metal.  Moreover, the anisotropy of the  electronic structure of Ce$M$In$_5$ is not large compared with high-$T_c$ cuprates; the anisotropy ratio of the Fermi velocity parallel and perpendicular to the $ab$-plane, $v_\parallel/v_\perp$, is more than 100 for high-$T_c$ cuprates, while it is $\sim$2 for Ce$M$In$_5$.     Therefore it seems highly unlikely that spin$-$charge separation in the 2D plane discussed in relation to cuprates is realized in CeRhIn$_5$ and CeCoIn$_5$.

In the Fermi liquid approach, a strong anisotropy of the transport scattering rate arising from the coupling to a singular bosonic mode, such as AF spin fluctuations,  is assumed \cite{pines,millis,kontani,rosch,hussey}.  Since the anomalous features in the magnetotransport properties are more pronounced when the system approaches the AF QCP, as shown by the $P$-dependence of $R_H$ in Figs.~ \ref{Hall2}~(a) and (b), it is natural to consider that the AF fluctuations play an essential role in determining the electron transport coefficients.  The $H$-dependence of $R_H$ provides strong support for this idea.  As indicated by resistivity\cite{QCP1}, heat capacity\cite{bian}, magnetization\cite{tayama}, and NMR relaxation rate\cite{NMR4} measurements, AF fluctuations are dramatically suppressed by the magnetic field,  and  recovery of  Fermi liquid behavior is observed above $\sim$5~T ($\sim\mu_0H_{c2}$) for $H\perp ab$-plane \cite{bian,QCP1}.    The insets of Figs.~\ref{Hall2} (a) and (b) show the $H$-dependence of $R_H={\rm d}\rho_{xy} /{\rm d} H$ of CeRhIn$_5$ and CeCoIn$_5$, respectively.   With increasing $H$, the amplitude of $R_H$ decreases rapidly and approaches $R_H$ of La$M$In$_5$.   At $\mu_0H$~=~7~T,  where the system exhibits  Fermi liquid behavior, $R_H$ is very close to $R_H$ of La$M$In$_5$.    Thus both the $P$- and $H$-dependence of $R_H$ imply that  AF fluctuations govern the electron transport phenomena below $T_{coh}$, particulary in the regime where  non-Fermi liquid behavior is observed.

\begin{figure*}[t]
\begin{center}
\includegraphics[width=10cm]{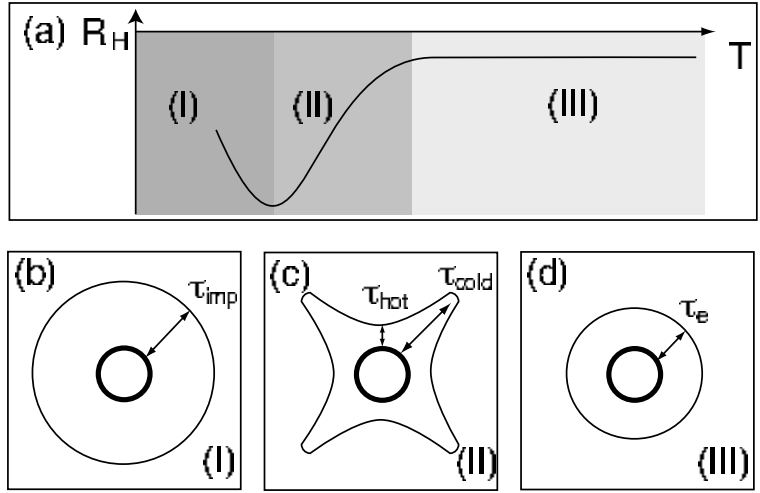}
\caption{(a) Schematic figure of the $T$-dependence of $R_H$ for CeCoIn$_5$ and CeRhIn$_5$.  There are three distinct $T$-regions: high-$T$-region (III), where $R_H$ is nearly $T$-independent; intermediate-$T$-region (II), where $R_H$ exhibits a rapid decrease with decreasing $T$; and low-$T$-region (I), where $R_H$ increases.  (b), (c), and (d) show the anisotropy of the  scattering time $\tau$($\phi$)  (thin solid lines) around the Fermi surface (solid lines) at region-(I), (II), and (III), respectively.  In region-(II), the scattering time is highly anisotropic,  possibly due to  AF spin fluctuations.  On the other hand,  the scattering time is isotropic in region-(I) where impurity scattering dominates and in region-(III) where  phonon and incoherent Kondo scattering determine the elastic scattering time $\tau_e$.    }
\label{HallT}
\end{center}
\end{figure*}

The central concept behind the Fermi liquid approach is the ``hot spot", a small area on the Fermi surface where the electron lifetime is unusually short.  Hot spots appear as a result of scattering of the electrons by  AF fluctuations.  They are located at  positions where the AF Brillouin zone intersects  the Fermi surface.     The presence of hot spots has been confirmed in high-$T_c$ cuprates.  It has  been reported in  CeIn$_3$ as well, which is a relative compound of Ce$M$In$_5$, by de Haas$-$van Alphen measurements \cite{ebihara}.   Since the hot spot area does not contribute to the electron transport, it reduces the effective carrier density, which results in enhancement of the amplitude of $R_H$ above $|1/ne|$.    However, we note that  the Hall effect and magnetoresistance in Ce$M$In$_5$ cannot be quantitatively accounted for by  Bloch$-$Boltzmann approaches assuming the presence of hot spots.  In fact, according to Ref. [\citen{pines}],  $R_H$ can be written as 
\begin{equation}
R_H=\frac{1}{en}\frac{1+r}{2\sqrt{r}},
\end{equation}  
where  $r$ is the anisotropy ratio of the scattering time at the hot and cold (non-hot) spots, $r \equiv \tau_{cold}/\tau_{hot}$.   Applying this theory to CeRhIn$_5$ and CeCoIn$_5$ yields an extremely large $r$:  $r\sim$ 10000 for CeRhIn$_5$ and $r\sim$ 2500 for CeCoIn$_5$.  Such an extremely large anisotropic scattering rate on the Fermi surface is unphysical and  inconsistent with some aspects of the experimental data.    For instance,  such a large $r$ would produce an extremely large magnetoresistance \cite{millis}, $\Delta \rho_{xx}/\rho_{xx}\sim 10r\tan^2 \Theta_H > 100$ at 1~T,  which is two orders of magnitude larger than the experimental results shown in Fig. \ref{MR1}. \\

\subsection{Anomalous enhancement of the Hall coefficient}

Figure~\ref{HallT} presents a schematic illustration of the $T$-dependence of $R_H$ for CeCoIn$_5$ and CeRhIn$_5$.  There are three distinct $T$-regions: high-$T$-region (III), where $R_H$ is nearly $T$-independent and $|R_H|\sim|1/ne|$; intermediate-$T$-region (II), where the amplitude of $R_H$ exhibits a rapid increase: and a low-$T$-region (I), where the amplitude of $R_H$ decreases.  Region-(I) is absent in the superconducting regime of CeRhIn$_5$ at $P>P_c$, as shown in Fig.~\ref{Hall2} (a).  In region-(III), phonon and incoherent Kondo scattering determine the elastic scattering time $\tau_e$.  We here address  the enhancement of the amplitude of $R_H$ in region-(II).

Given the failure of the Fermi liquid approach based on the Bloch$-$Boltzmann approximation to explain the Hall behavior in region-(II),  it has been suggested that the approach  should be susbtantially modified due to the backflow effect arising from strong electron correlation \cite{kontani}.  
The existence of  backflow is a  consequence of the charge conservation law, which is intimately related to the transport phenomena. Consequently, neglect of backflow often leads to various unphysical results. In general, backflow is not negligible in strongly correlated systems such as high-$T_c$ cuprates and heavy fermion systems.  The backflow is called the current vertex correction in microscopic Fermi liquid theory.

We here attempt to explain the non-Fermi liquid like behavior in the transport properties of CeRhIn$_5$ and CeCoIn$_5$ in accordance with  theory in which the backflow effect is taken into account.     Backflow is explained as follows.    According to the Landau-Fermi liquid theory, 
the energy of a quasiparticle with a wave-vector  {\boldmath $k$} is given by $\tilde{\varepsilon}_{\bf k}=\varepsilon_{\bf k}+\sum_{k'}f($ {\boldmath $k,k'$}$)\delta n_{k'}$, where $\varepsilon_{\bf k}=\hbar^2${\boldmath $k$}$^2/2m^{\ast}$, and $m^{\ast}$ is the renormalized mass.  $\delta n_{k}$ is the deviation of a quasiparticle distribution function from the ground state and  $f$({\boldmath $k,k'$}) is the Landau interaction function, which stems from electron$-$electron correlations.  
Therefore, whole the Fermi surface becomes unstable even if 
an excited quasiparticle exists.
Then, unstable Fermi surface induces excess current, which is called the backflow
 \cite{kontani-RV}.  

The total current in the collisionless limit ($\omega\tau\gg 1$) is given by 
\begin{eqnarray}
{\bf J}_{\bf k} &\equiv& ne{\bf \nabla}_{\bf k} \tilde{\varepsilon}_{\bf k}
 \nonumber \\
&=& ne{\bf v}_{\bf k}+ neN(0)\int_{\rm FS}dk'_\parallel 
f({\bf k},{\bf k'}){\bf v}_{\bf k'},
 \label{eqn:BS}
\end{eqnarray}
where ${\bf v}_{\bf k}=\hbar{\bf k}/m^*$ is the renormalized Fermi velocity and the second term represents the backflow.  $n$ is the density of electrions, and $N(0)$ is the density of states at the Fermi level.  
In a spherical system, ${\bf J}_{\bf k}$ is given by $\hbar{\bf k}/m$ due to the Galilean invariance of the system. Although $|{\bf J}_{\bf k}|$ is greater than $|{\bf v}_{\bf k}|$ due to backflow, {\boldmath $J$}$_k$ is parallel to {\boldmath $v$}$_{k}$. In anisotropic systems, however, ${\bf J}_{\bf k}$ is not always parallel to {\boldmath $v$}$_{k}$, which can modify the Hall coefficient significantly.
We stress that the dc conductivity is given by the total current in the hydrostatic limit ($\omega\tau\ll 1$), which is not as same as Eq. (\ref{eqn:BS}) \cite{kontani}.  Although the effect of backflow is more prominent in the hydrostatic limit, we use Eq. (\ref{eqn:BS}) hereafter for simplicity.  For the detailed argument, see Ref. [\citen{Yamada-textbook}].

\begin{figure*}[t]
\begin{center}
\includegraphics[width=14cm]{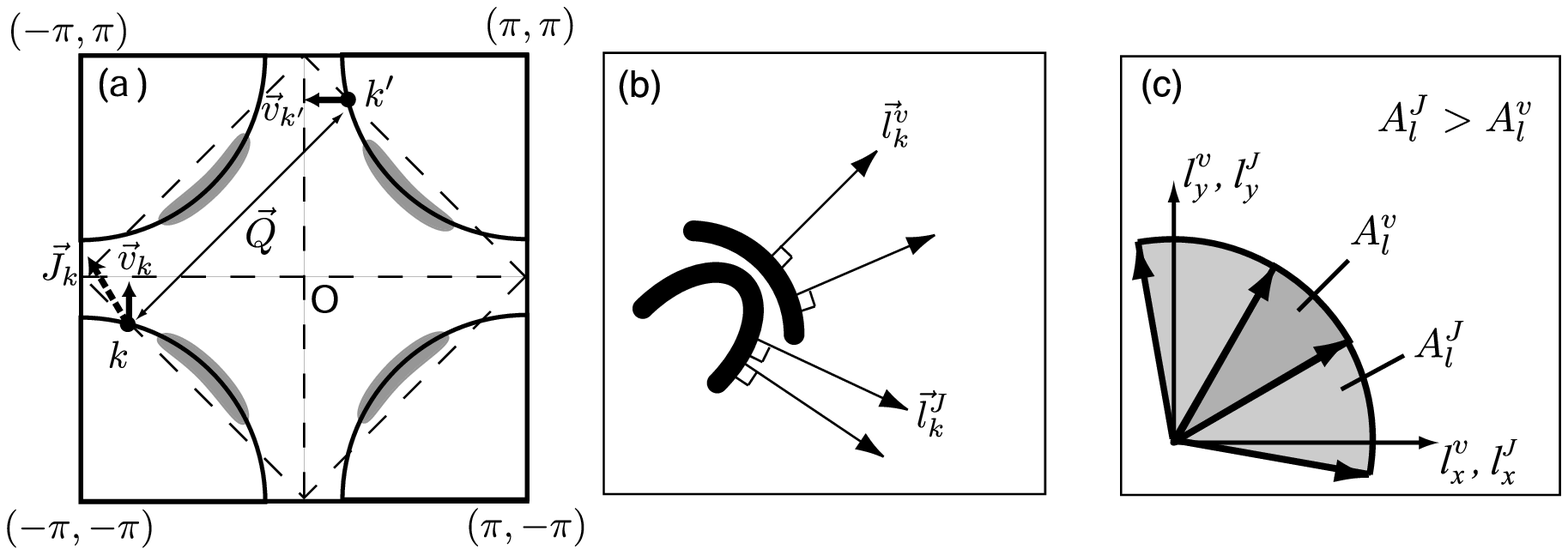}
\caption{(a) 2D Fermi surface.  In the presence of strong AF fluctuations,  {\boldmath $v$}$_k$  is strongly coupled  to {\boldmath $v$}$_{k'}$={\boldmath $v$}$_{k+Q}$ through a AF nesting vector {\boldmath $Q$}.   Shaded area represents cold spots. Soild arrows indicate {\boldmath $v_k$} and {\boldmath $v_{k'}$}.  Dashed arrow indicates {\boldmath $J$}$_k$. (b) Fermi surface with respect to {\boldmath $v$}$_k$ and {\boldmath $J$}$_k$ around a cold spot. The curvature of the Fermi surface can be strongly enhanced. (c) Swept area with respect to  {\boldmath $J_k$} and  that with respect to  {\boldmath $v_k$} around a cold spot.  $A_{\ell}^J$ can be strongly enhanced from $A_{\ell}^v$.    }
\label{Vertex}
\end{center}
\end{figure*}

Here we assume a 2D system for simplicity.  In the framework of the Bloch$-$Boltzmann approximation, the Hall conductivity is expressed as
\begin{equation}
\sigma_{xy}=(e^3/2\pi\hbar^2)A_{\ell},
\end{equation}
where $A_{\ell}^v$ is the area swept out by the vector $\mbox{\boldmath $\ell$}_k^v$ ($\parallel$  {\boldmath $v$}$_k$) as {\boldmath $k$} moves around the Fermi surface given as
\begin{equation}
A_{\ell}^v=\frac{{\bf B}}{2}\cdot \int d\mbox{\boldmath $\ell$}_k^v \times \mbox{\boldmath $\ell$}_k^v.
\end{equation}
Here, $\mbox{\boldmath $\ell$}_k^{v}$ ={\boldmath $v$}$_k\tau_k$ is the mean free path and {\boldmath $v$}$_k=\partial\epsilon_k/\partial(\hbar${\boldmath $k$}$)$ is the Fermi velocity at {\boldmath $k$}.\cite{ong3}   However, in the presence of the backflow effect, {\boldmath $J$}$_k$ is no longer parallel to {\boldmath $v$} and hence is not perpendicular to the Fermi surface.  Therefore,  backflow leads to a significant modification of the Hall coefficient and magnetoresistance from the Bloch$-$Boltzmann value.   Figures~\ref{Vertex} (a), (b), and (c) schematically show the backflow effect.   In the presence of strong AF fluctuation,  $f({\bf k},{\bf k'})$ is proportional to the spin susceptibility $\chi({\bf k}-{\bf k'})$. Therefore, according to Eq. (\ref{eqn:BS}), the backflow is proportional to {\boldmath $v$}$_{k+Q}$, where ${\bf Q}$ is the AF nesting vector. Then, the total current is approximately proportional to {\boldmath $v$}$_k +${\boldmath $v$}$_{k+Q}$.

We now consider the ``effective Fermi surface'' perpendicular to {\boldmath $J$}$_k$, which is illustrated in Fig.~\ref{Vertex}~(b).    When  {\boldmath $J$}$_k$ is not parallel to {\boldmath $v$}$_k$,  $A_{\ell}^v$ is replaced by the area swept out by $\mbox{\boldmath $\ell$}_k^{J}$ ={\boldmath $J$}$_k\tau_k/ne$,  as shown in Fig.~\ref{Vertex}~(c),  
\begin{equation}
A_{\ell}^J=\frac{{\bf B}}{2}\cdot \int d\mbox{\boldmath $\ell$}_k^J \times \mbox{\boldmath $\ell$}_k^J.
\end{equation}
Around the cold spot, the curvature of the effective Fermi surface is strongly enhanced.  Then, the swept area with respect to  {\boldmath $J_k$} can be strongly enhanced, compared with that with respect to  {\boldmath $v_k$} ($A_{\ell}^J \gg A_{\ell}^v$).   Thus the backflow effect leads to a significant enhancement of the Hall coefficient.

We next discuss the transport coefficient in the presence of the backflow effect in more detail.   In the presence of strong anisotropy of the scattering rate at the Fermi surface, various transport properties are determined by $\tau_{cold}$.  According to Ref.~[\citen{kontani}], transport properties under a magnetic field are also governed  by the AF correlation length $\xi_{AF}$ in the presence of the backflow effect.   In this case, the zero-field diagonal conductivity $\sigma_{xx}(0)$, Hall conductivity $\sigma_{xy}$ \cite{kontani}, and magnetoconductivity $\Delta\sigma_{xx}(H)\equiv \sigma_{xx}(H)-\sigma_{xx}(0)$ 
\cite{kontani-MR}
are given as
\begin{equation}
	\sigma_{xx}(0)\sim \tau_{cold},
\end{equation}
\begin{equation}
	\sigma_{xy} \sim \xi_{AF}^2\tau_{cold}^2H,
\end{equation}
and
\begin{equation}
	\Delta\sigma_{xx}\sim \xi_{AF}^4\tau_{cold}^3H^{2}.
\end{equation}
Here we note that the non-linearity of $\sigma_{xy}$ with respect to $H$ is not due to the higher order terms with respect to $H\tau$ because $\omega_c\tau \ll 1$. ($\omega_c= e\mu_0H/m$ being the cyclotron frequency.) Near the AF QCP, $\xi_{AF}$ depends on $T$ as $\xi_{AF}^2 \propto 1/(T+\theta)$\cite{comment}, where $\theta$ is the Weiss temperature.   Moreover, according to AF spin fluctuation theory, $\tau_{cold}$ is nearly inversely proportional to $T$: $\tau_{cold}\propto 1/T$.\cite{kontani}    We then obtain the temperature dependence of the resistivity $\rho_{xx}=\sigma_{xx}^{-1}$, Hall coefficient $R_{H}=\sigma_{xy}/\sigma_{xx}^2\mu_0H$,  and  Hall angle as
\begin{equation}
	\rho_{xx} \propto\tau_{cold}^{-1} \propto T,
\label{eqn:rho}
\end{equation}
\begin{equation}
	R_{H} \propto \xi_{AF}^2 \propto \frac{1}{T},
\label{eqn:RH}
\end{equation}

\begin{equation}
	\cot\Theta_H \propto T^2.
\end{equation}
These well reproduce the observed $T$-dependence of $\rho_{xx}$, $R_H$, and $\cot \Theta_H$  of CeRhIn$_5$ and CeCoIn$_5$ shown in Figs.~\ref{DCres} (a) and (b) and Figs~\ref{Hall5} (a), (b), and (c).    We note that a recent calculation based on the band structure of CeCoIn$_5$ is quantitatively compatible with the enhancement of $R_H$ below $T_{coh}$\cite{onari}.

Enhancement of the amplitude of $R_H$ at low temperatures has been reported for other quasi-2D strongly correlated systems, $\kappa$-(BEDT-TTF)$_2$Cu(NCS)$_2$ \cite{katayama,sushko,kontani-kappa} and  $R_{2-x}$Bi$_x$Ru$_2$O$_7$ ($R$~=~Sm and Eu)\cite{satom}.  In addition, enhancement has also been reported for the 3D strongly corrrelated system V$_{2-y}$O$_3$\cite{VO}.   It has been shown that backflow plays an important role even in a 3D system, irrespective of the fact that the backflow vanishes in the $d=\infty$ limit. Therefore, the enhancement of the Hall coefficient in these three strongly correlated systems may relate to the backflow mechanism, though the  $T$-dependence of the transport coefficients seem to be very different from those of Ce$M$In$_5$ and high-$T_c$ cuparates.

We also note that in addition to these dc Hall effect, recent measurements of the high frequcncy ac Hall effect for high-$T_c$ cupartes  have reported  anomalous $T$-and $\omega$- dependence which exhibit a striking deviation from the Fermi liquid behavior \cite{drew1,drew2}.  This behavior has been discussed in the light of the backflow effect \cite{kontani-ACHall}.  \\

\subsection{Modified Kohler's rule}

By definition, the magnetoresistance is given by
\begin{equation}
	\frac{\Delta\rho_{xx}(H)}{\rho_{xx}(0)}=- \frac{\Delta\sigma_{xx}(H)}{\sigma_{xx}(0)}-\left (\frac{\sigma_{xy}(H)}{\sigma_{xx}(0)}\right )^2.   
\end{equation}
Using the relation $\Delta\sigma_{xx}(H)/\sigma_{xy}(H)^2\sim \tau_{cold}^{-1}$, given by Eqs.~(11) and (12), the magnetoresistance is obtained as
\begin{equation}
	\frac{\Delta\rho_{xx}(H)}{\rho_{xx}(0)}=(\tan\Theta_H)^2\cdot\left(\frac{\sigma_{xx}(H)}{\sigma_{xx}(0)}\right)^{2}\cdot(C-1),
\end{equation}
where $C$ is a constant and is much larger than unity for Ce$M$In$_5$.   Since $\sigma_{xx}(H)/\sigma_{xx}(0)\simeq 1$ at low fields, $\Delta\rho_{xx}(H)/\rho_{xx}(0)$ should be well scaled by $\tan^2 \Theta_H$.   Thus Eq.~(6) holds for a theory in which the backflow effect is taken into account for the Fermi liquid ground state.  

Since 
\begin{equation}
\Delta\rho_{xx}(H)/\rho_{xx}(0)\sim \xi_{AF}^{4}\tau_{cold}^{2}H^2,
\end{equation}
 the $P$- and $H$-dependence of the magnetoresisitance is governed by the $P$- and $H$-dependence of $\xi_{AF}$.  As shown in Figs.~\ref{MR1} (a), (b), and (c), the magnetoresistance shows an anomalous concave downward curvature at $P<P^{\ast}$ (see Figs.~1~(a) and (b)), while such a behavior is not observed at $P>P^{\ast}$.  Since the AF fluctuations are dramatically suppressed by  magnetic fields at $P<P^{\ast}$\cite{bian,QCP1,tayama,NMR4}, as discussed in \S4.1,  $\xi_{AF}$ is rapidly reduced by magnetic fields.  On the other hand, at $P>P^{\ast}$ where the AF fluctuation is already suppressed in zero field, $\xi_{AF}$ is insensitive to the magnetic field.   Thus the $P$- and $H$-dependence of the magnetoresistance in CeCoIn$_5$ and CeRhIn$_5$ reinforce the conclusion that AF fluctuations near the QCP play an essential role in determining  the transport properties.\\

\subsection{Impurity and pseudogap effects on the Hall effect}

Having established the evidence for the importance of AF fluctuations on the electron transport properties, we next discuss the influence of impurities and the pseudogap on the Hall effect at low temperatures.

The amplitude of $R_H$ is reduced at low emperatures in CeCoIn$_5$ and CeRhIn$_5$ at $P<P_c$, while such a reduction does not occur in CeRhIn$_5$ at $P>P_c$ (see Figs.~\ref{Hall2}~(a) and (b) and Fig.~\ref{Hall3}).   We note that there are two possible origins for this: impurities and pseudogap formation.

We first discuss why region-(I) illustrated in Fig.~\ref{HallT} is present in CeCoIn$_5$ while it is absent in CeRhIn$_5$ at $P>P_c$.   We note that this can be accounted for by considering the impurity scattering effect.  The scattering time  $\tau(\phi)$  can be taken to be the sum of an isotropic impurity part and  anisotropic inelastic part: 
\begin{equation}
	\tau(\phi)=\frac{1}{\tau_{imp}^{-1}+\tau_{inel}(\phi)^{-1}}. 
\end{equation}
In  region-(II) where inelastic scattering due to AF fluctuation dominates,  backflow becomes important.  However, as the temperature is lowered, isotropic impurity scattering becomes important ($\tau_{imp}^{-1}>\tau_{inel}^{-1}$).  As a result, the influence of  AF fluctuation on the Hall effect is reduced and  the scattering become isotropic as shown in Fig.~\ref{HallT}~(b). In this situation, the backflow effect is  greatly reduced\cite{kontani} and $R_H$ is expected to approach the Fermi liquid value.   The fact that $R_H$ at $P$ = 2.51~GPa in CeCoIn$_5$ approaches $R_H$ of LaCoIn$_5$ seems to support this. (At $P$ = 0 and 1.04~GPa, $R_H$ does not recover to $R_H^{La}$ in region-(I), because the system undergoes a superconducting transition.)  To obtain more insight into the impurity effect, we compare the resistivity values at the temperatures $T_m$ where the amplitude of $R_H$ begin to decrease with decreasing $T$.   The values of $T_m$ at $P$~=~0, 1.04, and 2.51~GPa obtained from Fig.~\ref{Hall2}~(b) are $\sim$4~K, 6~K, and 20~K, respectively.  The values of $\rho_{xx}^{mag}$ at $T_m$ are $\sim$ 6.6, 4.9, and 6.1 $\mu\Omega$cm at $P$=0, 1.04, and 2.51~GPa, respectively.  Thus $\rho_{xx}^{mag}$ at $T_m$ have similar values, which are nearly $P$-independent.  This can be seen by the hatched region in Fig. \ref{DCres}~(c).

Moreover, the impurity scenario naturally explains the absence of region-(I) in CeRhIn$_5$ at $P>P_c$.  According to recent NMR $T_1^{-1}$ measurements in the superconducting state, the impurity concentration in CeRhIn$_5$ is lower than that in CeCoIn$_5$, indicating that the impurity scattering in CeRhIn$_5$ is expected to be less important than in CeCoIn$_5$.  Based on this finding, we conclude that {\it impurity scattering seriously influences the low temperature Hall effect.}

On the other hand, the origin of the reduction of the amplitude of $R_H$ observed in the AF metallic regime of CeRhIn$_5$ seems to be more complicated.   In fact, in contrast to CeCoIn$_5$, $\rho_{xx}$ at different pressures have very different values at $T_m$, as shown by the dashed arrows in the inset of Fig.~\ref{Hall3}.   We thus speculate that the reduction of $\xi_{AF}$ due to the formation of a gap in the magnetic exciation is a primary cause of the reduction in the amplitude of $R_H$, because of the following reason.

\begin{figure}[t]
\begin{center}
\includegraphics[width=7.5cm]{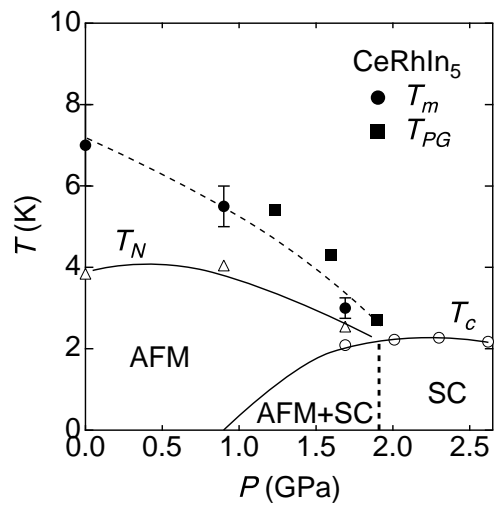}
\caption{ Pressure dependence of the temperature at which the amplitude of $R_H$ exhibits a maximum, $T_m$ (filled circle), pthe seudogap temperature determined by NMR relaxation rate $T_{PG}$ (filled squares), and the N\'eel temperature $T_N$ in the AF metallic regime of CeRhIn$_5$ ($P<P_c$).   }
\label{PG}
\end{center}
\end{figure}

Recentl nuclear quadrupole resonance measurements indicate that the NMR  $T_1^{-1}$ exhibits a peak at $T_{PG}$ above $T_N$ \cite{kawasaki}, indicating a reduction of the AF correlation below $T_{PG}$.    This behavior has been discussed in terms of gap formation in the spin excitation or a possible formation of a pseudogap in analogy to high-$T_c$ cuprates.  In Fig.~\ref{PG},  the pressure dependence of $T_{PG}$ reported in Ref.~[\citen{kawasaki}] is plotted.  As seen in Fig.~\ref{PG}, $T_m$ is  located close to $T_{PG}$.  This implies a close relationship between the Hall effect and pseudogap formation.  In this situation, the anisotropy of the inelastic scattering time due to  AF spin fluctuations is expected to be seriously reduced by the suppression of  magnetic scatterings, which gives rise to the situation illustrated in Fig.~\ref{HallT}~(b).   More detailed investigation is necessary to clarify this issue.  \\

\subsection{Hall effect and QCP}  

As discussed in $\S$1, the evolution of the Fermi surface  when crossing the QCP  is an important issue in strongly correlated electron systems.    Two contrary views have been proposed.    Theories based on  SDW predict a continuous change of the Fermi surface \cite{hertz, millis2},  while theories based on the breakdown of composite Fermions predict a jump of  the Fermi surface volume at the QCP\cite{coleman}.   As pointed out in Ref.~[\citen{coleman}],    Hall measurements that can probe the Fermi surface are important for clarifying this controversial issue.   According to the SDW picture,  the Hall coefficient is expected to change continuously when crossing the QCP, while according to the picture of the breakdown of composite Fermions,  the Hall coefficient is expected to jump at the QCP \cite{coleman}.

As shown in Fig.~\ref{Hall4},   $|R_H^{\max}|$  for CeRhIn$_5$ at $\mu_0H$~=~0 increases rapidly as  $P_c$ is approached from either the AF metallic ($P<P_c$) or superconducting ($P>P_c$) regimes without exhibiting a jump at $P_c$.   At $\mu_0H$~=~7~T, the enhancement of $|R_H^{\max}|$ is strongly reduced and exhibits a broad maximum at a pressure slightly higher than $P_c$.  Recently it has been reported that the QCP slightly increases from $P_c$ under a magnetic field and is located at $\sim$ 2.1 GPa at $\mu_0H$~=~7~T \cite{park}.  Thus  $|R_H^{\max}|$ in CeRhIn$_5$ appears to exhibit a maximum  at or in the vicinity of the QCP without showing a jump.   These results lead us to consider that a transition from  AF metal with a small Fermi surface to a Fermi liquid with a large Fermi surface at the QCP suggested in Ref.~[\citen{coleman}] is  unlikely to occur in CeRhIn$_5$.  This Hall behavior in CeRhIn$_5$ is in striking contrast to the recent Hall measurements on YbRh$_2$Si$_2$, in which the Hall coefficient exhibits a discontinuous jump at the QCP \cite{paschen}.\\

\subsection{Multiband effects}  

We here stress that the multiband effect is very unlikely to be an origin of the observed magnetotransport anomalies in Ce$M$In$_5$.

Real heavy fermion systems possess multiple Fermi surfaces with different sizes. According to dHvA experiments, quasiparticle masses on larger or smaller Fermi surfaces are, respectively,  heavier or lighter in general. One may consider that lighter quasiparticles on smaller Fermi surfaces would be the main contribution to the transport phenomena, irrespective of the small area of the Fermi surface. However, it is well known that the Kadowaki-Woods relation, 
$A\gamma^{-1}=10^{-5}/\frac12 N(N-1) ~\mu\Omega$cm(mol$\cdot$K/mJ)$^2$,
holds for many heavy fermion superconductors, where $A$ is the coefficient of the $T^2$ term in the resistivity, $\gamma$ is the $T$-linear specific heat coefficient, and  $N$ is the $f$-orbital degeneracy \cite{tsujii}. This fact suggests that the transport phenomena are governed by larger Fermi surfaces with heavier masses in many compounds.

First of all,  that Kohler's rule holds for LaRhIn$_5$ over four orders of magnitude of $\Delta\rho_{xx}/\rho_{xx}$ implies that the multiband effect is not important for the magnetotransport in Ce$M$In$_5$.  Moreover, as we have discussed, the anomalous behaviors of $\rho_{xx}$, $R_H$, and $\Delta\rho_{xx}/\rho_{xx}$ in CeMIn$_5$ are satisfactorily scaled by a single parameter $\xi_{AF}$ for a wide rang of  $T$, $H$, and $P$. This fact strongly suggests that the transport phenomena are caused by a single large Fermi surface composed of heavy quasiparticles. However, CeMIn$_5$ is a multiband system; the 13th and 14th bands give small and large hole-like Fermi surfaces, respectively, while the 15th and 16th bands give large and small electron-like Fermi surfaces, respectively \cite{maehira}.  One might consider that the small electron-like Fermi surfaces would be the origin of the huge negative $R_H$ in CeMIn$_5$ near the AF QCP. However, this possibility is ruled out by the following argument.

We here discuss the possible multiband effect on the Hall coefficient by considering a two-band model ($\alpha$, $\beta$).  We express the Hall coefficient and the conductivity in each band as ($R_H^\alpha=1/n^\alpha e$, $\sigma_{xx}^\alpha$) or ($R_H^\alpha=1/n^\beta e$, $\sigma_{xx}^\alpha$). Here we assume $R_H^\alpha > R_H^\beta >0$. Then, the Hall coefficient of this system is 
\begin{eqnarray}
R_H= R_H^\alpha \frac{x^2+ R_H^\beta/R_H^\alpha}{(x+1)^2},
\end{eqnarray}
where $x=\sigma_{xx}^\alpha/\sigma_{xx}^\beta$. We can show that $R_H^{\rm max}= R_H^\alpha$ at $x=\infty$,  and $R_H^{\rm min}= R_H^\beta \frac{R_H^\beta/R_H^\alpha+1}  {(R_H^\beta/R_H^\alpha+1)^2} \lesssim R_H^\beta$ at $x=R_H^\beta/R_H^\alpha$. It is easy to see that
\begin{eqnarray}
R_H &\sim& R_H^\alpha \ \ \ {\rm for} \ \ \ 
 \sigma_{xx}^\alpha \gtrsim \sigma_{xx}^\beta, \label{eqn:RH-large} \\
R_H &\sim& R_H^\beta \ \ \ {\rm for} \ \ \ 
 \sigma_{xx}^\alpha \ll \sigma_{xx}^\beta. \label{eqn:RH-small}
\end{eqnarray}
Therefore, the assumption $R_H^\alpha/R_H^\beta =n^\beta/n^\alpha=30-50$ is necessary to explain the experimental results of the Hall effect in CeMIn$_5$. According to the band calculation, $n$(15th band)$/n$(16th band)$=0.626/0.030=21$. In CeCoIn$_5$ at ambient pressure, Eq.~(\ref{eqn:RH-large}) should be satisfied for $T \gtrsim T_c$, whereas Eq.~(\ref{eqn:RH-small}) has to be realized for $T \sim T_{coh}$.  That is, the most conductive band changes from $\alpha$ to $\beta$ with increasing temperature.  At $P$~=~2.51~GPa, on the other hand, Eq.~(\ref{eqn:RH-small}) should be satisfied below $T_{coh}$.  In this case, the most conductive band is always $\beta$.  However, we see in Fig.~\ref{DCres}~(c) that $\rho_{xx}^{\rm mag}$ at $P$~=~0 and 2.51~GPa shows very similar $T/T_{coh}$-dependences.  This experimental result apparently contradicts the above mentioned scenario. Moreover, the present two-band model cannot explain the enhancement of the magnetoresistance and the Nernst coefficient below $T_{coh}$, because both are independent of the band filling in the Bloch$-$Boltzmann approximation.\\

\subsection{Thermoelectric effects}  

In addition to the non-Fermi liquid behavior of the electron transport properties, the thermoelectric effects, such as Nernst coefficient and thermoelectric power,  have been reported to be anomalous in high-$T_c$ cuprates.  Especially,  a giant enhancement of the Nernst coefficient $\nu_H$ three orders of magnitude larger than that expected in conventional metals has been reported in the normal state of high-$T_c$ cuprates in the underdoped and optimally doped regimes \cite{nernst}.  Recently, it has been reported that the amplitude of  $\nu_H$ in the normal state of CeCoIn$_5$ at ambient pressure is also strikingly enhanced by more than three orders of magnitude, compared with conventional metals\cite{bel}. The sign of $\nu_H$ is positive in cuprates while it is negative in CeCoIn$_5$.

The origin of the giant Nernst signal is controversial.  On the one hand, the origin has been discussed in the light of vortex-like excitations in high-$T_c$ cuprates.  On the other hand, it has been suggested that the origin can be explained in a unified way in the framework of the backflow effect \cite{kontani-N}. Whether the giant Nernst coefficients in high-$T_c$ cuprates and CeCoIn$_5$ have the same origin should be scrutinized.  In CeCoIn$_5$, the giant Nernst effect gradually fades with the application of a magnetic field.   This phenomenon, like the anomalously large Hall coefficient,  may reflect a large energy-dependence of the scattering time and  highly anisotropic backflow due to the presence of AF fluctuations.    A study of the pressure dependence of the Nernst effect in Ce$M$In$_5$ is necessary to clarify this issue.  \\

\section{Conclusions}

In summary, we have performed a systematic study of the magnetotransport properties of quasi 2D heavy fermion Ce$M$In$_5$ ($M$:Rh and Co) under pressure, ranging from the AF metal state to the Fermi liquid state through the non-Fermi liquid state.

In Ce$M$In$_5$, the amplitude of the Hall coefficient is dramatically enhanced with decreasing temperature and attains a value much larger than $|1/ne|$ at low temperatures.    Furthermore, the magnetoresistance exhibits striking violation of Kohler's rule.  These findings are in shrap contrast to the bahavior of La$M$In$_5$ with a similar electronic structure.  We found that the dc-resistivity is proportional to $T$, the cotangent of the Hall angle $\cot \Theta_H$ varies as $T^2$, and the magnetoresistance is well scaled by  the tangent of the Hall angle as $\Delta \rho_{xx}/\rho_{xx}\propto \tan^2\Theta_H$.   These non-Fermi liquid behavior observed in the electron  transport are remarkably pronounced with approaching the AF QCP.  The $T$-, $H$-, and $P$-dependence of the dc-resistivity, Hall coefficient and magnetoresistance  definitely indicate that AF fluctuations play a crucial role in determining the non-Fermi liquid behavior of the electron transport phenomena.

We have shown that the anomalous electron transport properties can be accounted for in terms of a recent theory in which hot and cold spots around the Fermi surface and the backflow effect are taken into account; all of the anomalous behavior of dc-resistivity,  Hall effect, and magnetoresistance are  well scaled by a single parameter, the AF correlation length $\xi_{AF} \propto 1/\sqrt{T+\theta}$, for a wide range of the parameters, $T$, $H$, and $P$.

Several anomalous transport properties observed in Ce$M$In$_5$ bear a striking resemblance to the normal-state properties of high-$T_c$ cuprates, indicating  universal transport properties of strongly correlated quasi 2D electron systems in the presence of strong AF fluctuations.  Thus the present results may also prove relevant for  the debate on the anomalous normal-state properties of high-$T_c$ cuprates, holding the promise of bridging our understanding of heavy fermion systems and cuprates.  \\

\section*{Acknowledgment}

We thank  M.~Sato and K.~Yamada for stimulating discussions.  This work was partly supported by a Grant-in-Aid for Scientific Reserch from the Ministry of Education, Culture, Sports, Science and Technology.  One of the authors (Y.~N.) was supported by the Research Fellowships of the Japan Society for the Promotion of Science for Young Scientists.\\


\begin{thebibliography}{99}

 \bibitem{hall}C.~M.~Hurd: {\it The Hall Effect in Metal and Alloys} (Plenum, New York, 1972).
\bibitem{pippard}A.~B.~Pippard: {\it Magnetoresistance in Metals} (Cambridge University Press, 1989).
 \bibitem{moriya} T.~Moriya, and T.~Takimoto: J. Phys. Soc . Jpn, \textbf{64} (1995) 960. 
 \bibitem{coleman} P.~Coleman, C.~P\'epin, Q.~Si and R.~Ramazashvili: J.Phys.: Condens. Matter \textbf{13}  (2001) R723. 
 \bibitem{sachdev}S. Sachdev: {\it Quantum Phase Transitions},  (Cambridge University Press, Cambridge UK 1999).
\bibitem{sigrist}M.~Sigrist and K.~Ueda: Rev. Mod. Phys. \textbf{63}  (1991) 239.
\bibitem{thalmeier}P.~Thalmeier, and G.~Zwicknagl: {\it Handbook on the Physics and Chemistry of Rare Earths} (Elsevier, Amsterdam,2005) Vol.34, 135.
\bibitem{mineev}V.P.~Mineev and K.V.Samokhin: {\it Introduction to Unconventional Superconductivity} (Gordon and Breach Science Publishers, 1999).
\bibitem{ong1}see, e.g., N.~P.~Ong: {\it Physical Properties of High Temperature Superconductors}, edited by D.M.~Ginsberg (World Scientific, Singapore, 1992), Vol.2.
\bibitem{takeda}J.~Takeda, N.~Nishikawa, and M.~Sato: Physica C {\bf 231} (1994) 293.
\bibitem{ando}Y.~Ando,Y.~Kurita, S.~Komiya, S.~Ono, and K.~Segawa: Phys. Rev. Lett. {\bf 92} (2004) 197001.
\bibitem{chien} T.R.~Chien, Z.Z.~Wang, and N.P.~Ong, Phys. Rev. Lett: {\bf 67} (1991) 2088.
\bibitem{anderson}P.W.~Anderson: Phys. Rev. Lett. {\bf 67} (1991) 2092.
\bibitem{harris}J.M.~Harris, Y.F.~Yan, P.~Matl, N.P.~Ong, P.W.~Anderson, T.~Kimra and K.~Kitazawa: Phys. Rev. Lett. {\bf 75} (1995) 1391.
\bibitem{nernst}Z.A.~Xu, N.P.~Ong, Y.~Wang, T.~ Kakeshita, and S.~Uchida: Nature 406 (2000) 486.
\bibitem{pines} B.P.~Stojkovi{\' c} and D.~Pines: Phys. Rev. B {\bf 55}  (1997) 8576.
\bibitem{millis}L.B.~Ioffe and A.J.~Millis: Phys. Rev. B {\bf 58}  (1998) 11631.
\bibitem{kontani}H.~Kontani, K.~Kanki, and K.~Ueda: Phys. Rev. B {\bf 59} (1999) 14723, and K.~Kanki and H.~Kontani: J. Phys. Soc. Jpn,  {\bf 68} (1999) 1614.
\bibitem{rosch}A.~Rosch: Phys. Rev. B {\bf 62} (2000) 4945.
\bibitem{hussey}N.E.~Hussey: Eur. Phys. J. B {\bf 31} (2003) 495, and references therein.
\bibitem{ong2}N.P.~Ong and P.W.~Anderson: Phys. Rev. Lett. {\bf 78} (1997) 977.
\bibitem{pet1}C.~Petrovic, P.G.~Pagliuso, M.F.~Hundley, R.~Movshovich, J.L.~Sarrao,J.D.~Thompson, Z.~Fisk and P.~Monthoux: J.\ Phys. Condens. Matter {\bf 13}  (2001) L337.
\bibitem{hegger}H.~Hegger, C.~Petrovic, E.G.~Moshopoulou, M.F.~Hundley, J.L.~Sarrao, Z.~Fisk, and J.D.~Thompson: Phys. Rev. Lett. {\bf 84} (2000) 4986. 
\bibitem{pet2}C.~Petrovic, R.~Movshovich, M.~Jaime, P.G.~Pagliuso, M.F.~Hundley, J.L.~Sarrao, Z.~Fisk, and J.D.~Thompson: Europhys. Lett. {\bf 53} (2001) 354.
\bibitem{settaisan}R.~Settai, H.~Shishido, S.~Ikeda, Y.~Murakawa, M.~Nakashima, D.~Aoki, Y.~Haga, H.~Harima and Y. Onuki: J. Phys.: Condens. Matter {\bf 13} (2001) L627.
\bibitem{shishido}H.~Shishido, R.~Settai, D.~Aoki, S.~Ikeda, H.~Nakawaki, N.~Nakamura, T.~Iizuka, Y.~Inada, K.~Sugiyama, T.~Takeuchi, K.~Kindo, T.C.~Kobayashi, Y.~Haga, H.~Harima, Y.~Aoki, T.~Namiki, H.~Sato and Y.~Onuki: J. Phys. Soc. Jpn {\bf 71} (2002) 162.
\bibitem{NMR}G.-q.~Zeng, K.~Tanabe, T.~Mito, S.~Kawasaki, Y.~Kitaoka, D.~Aoki, Y.~Haga and Y.~Onuki: Phys. Rev. Lett. {\bf 86} (2001) 4664.
\bibitem{NMR2}Y.~Kawasaki, S.~Kawasaki, M.~Yashima, T.~Mito,G.-q~Zheng, Y.~Kitaoka, H.~Shishido, R.~Settai, Y.~Haga and Y.~Onuki: J. Phys. Soc.Jpn. {\bf 72} (2003) 2308.
\bibitem{sid} V. A.~Sidorov, M.~Nicklas, P.G.~Pagliuso, J.L.~Sarrao, Y.~Bang, A.V.~Balatsky and J.D.~Thompson: Phys. Rev. Lett. {\bf 89} (2002) 157004.
\bibitem{kneb} G.~Knebel, M-A.~M\'{e}asson, B.~ Salce, D.~Aoki, D.~Braithwaite, J.P.~Brison and J.~Flouquet: J. Phys.: Condens. Matter {\bf 16} (2004) 8905.
\bibitem{fisher}R. A.~Fisher, F.~Bouquet, N.E.~Phillips, M.F.~Hundley, P.G.~Pagliuso, J.L.~Sarrao, Z.~Fisk, and J.D.~Thompson: Phys. Rev. B {\bf 65} (2002) 224509.
\bibitem{ikeda}S.~Ikeda, H.~Shishido, M.~Nakashima, R.~Settai, D.~Aoki, Y.~Haga, H.~Harima, Y.~Aoki, T.~Namiki, H.~Sato and Y.~Onuki: J. Phys. Soc. Jpn. {\bf 70} (2001) 2248.
\bibitem{bian} A.~Bianch, R.~Movshovich, I.~Vekhter, P.G.~Pagliuso and J.L.~Sarrao: Phys. Rev. Lett. {\bf 91} (2003) 257001.
\bibitem{tayama} T. Tayama, A. Harita, T. Sakakibara, Y. Haga, H. Shishido, R. Settai and Y. Onuki: Phys. Rev. B {\bf 65} (2002) 180504.
\bibitem{NMR3}Y.~Kohori, Y.~Yamato, Y.~Iwamoto, T.~KOhara, E.D.~Bauer, M.B.~Maple and J.L.~Sarro: Phys. Rev. B {\bf 64},(2001) 134526.
\bibitem{NMR4}M.~Yashima, S.~Kawasaki, Y.~Kawasaki, G.-q.~Zheng, Y.~Kitaoka, H.~Shishido, R.~Settai, Y.~Haga and Y.~Onuki:J. Phys. Soc. Jpn. {\bf 73} (2004) 2073.
\bibitem{bel}R.~Bel, K.~Behnia, Y.~Nakajima, K.~Izawa, Y.~Matsuda, H.~Shishido, R.~Settai and Y.~Onuki: Phys. Rev. Lett. {\bf 92} (2004) 217002.
\bibitem{mov} R.~Movshovich, M.~Jaime, J.D.~Thompson, C.~Petrovic, Z.~Fisk, P.G.~Pagliuso and J.L.~Sarrao: Phys.  Rev. Lett. {\bf 86} (2001) 5152.
\bibitem{aoki} H.~Aoki, T.~Sakakibara, H.~Shishido, R.~Settai, Y.~Onuki, P.~Miranovic and K.~Machida: J Phys.: Condens Matter {\bf 16} (2004)  L13.
\bibitem{izawa} K.~Izawa, H.~Yamaguchi, Y.~Matsuda, H.~Shishido, R.~Settai and Y.~Onuki: Phys. Rev. Lett. {\bf 87} (2001) 057002.
\bibitem{pene1}R.J.~Ormeno, A.~Sibley, C.E.~Gough, S.~Sebastian and I.R.~Fishier: Phys. Rev. Lett. {\bf 88} (2002) 47005.
\bibitem{pene2}E.E.M.~Chia, D.J.~Van~Harlingen, M.~B.~Salamon, B.D.~Yanoff, I.~Bonalde, and J.L.~Sarrao: Phys. Rev. B {\bf 67} (2003) 014527. 
\bibitem{wei}  P.M.C.~Rourke, M.A.~Tanatar,C.S.~Turel, J.~Berdeklis, C.~Petrovic, and J.Y.T.~Wei: Phys. Rev. Lett. {\bf 94} (2005) 107005 (2005).
\bibitem{green}W.K.~Park, L.~H.~Greene, J.~L.~Sarrao, and J.D.~Thompson: cond-mat/0606535
\bibitem{vekhter}A.~Vorontsov and I.~Vekhter: Phys. Rv. Lett. {\bf 96} (2006) 237001.
\bibitem{FFLO}H.A.~Radovan, N.A.~Fortune, T.P.~Murphy, S.T.~Hannahs, E.C.~Palm, S.W.~Tozer and D.~Hall: Nature {\bf 425} (2003) 51,  A.~Bianchi, R.~Movshovich, C.~Capan, P.G.~Pagliuso, J.L.~Sarrao: Phys. Rev. Lett. {\bf 91} (2003) 187004,  T.~Watanabe, Y.~Kasahara, K.~Izawa, T.~Sakakibara, Y.~Matsuda, C.J.~van der Beek, T.~Hanaguri, H.~Shishido, R.~Settai and Y.~Onuki: Phys. Rev. B {\bf 70} (2004) 020506, K.~Kakuyanagi, M.~Saitoh, K.~Kumagai, S.~Takashima, M.~Nohara, H.~Takagi and  Y.~Matsuda:  Phys. Rev. Lett. {\bf 94} (2005) 047602.
\bibitem{kasahara}Y.~Kasahara, Y.~Nakajima, K.~Izawa, Y.~Matsuda, K.~Behnia, H.~Shishido, R.~Settai, and Y.~Onuki: Phys. Rev. B {\bf 72} (2005) 214515.
\bibitem{fert}A.~Fert, and P.M.~Levy: Phys. Rev. B \textbf{36}  (1987) 1907.
\bibitem{coleman2}P.~Coleman, P.W.~Anderson, and T.V.~Ramakrishman: Phys. Rev. Lett. \textbf{ 55}  (1985) 414.
\bibitem{konyam}H.~Kontani and K.~Yamada: J. Phys. Soc. Jpn. {\bf 63} (1994) 2627.
\bibitem{nakajima1}Y.~Nakajima, K.~Izawa, Y.~Matsuda, S.~Uji, T.~Terashima, H.~Shishido, R.~Settai, Y.~Onuki, and H.~Kontani: J. Phys. Soc. Jpn {\bf 73} (2004) 5.
\bibitem{nakajima2}Y.~Nakajima, K.~Izawa, Y.~Matsuda, K.~Behnia, H.~Kontani, M.~Hedo, Y.~Uwatoko, T.~Matsumoto, H.~Shishido, R.~Settai, and Y.~Onuki: J.Phys. Soc. Jpn {\bf 75} (2006) 023705.
\bibitem{hertz}J.A.~Hertz: Phys. Rev. B {\bf 14} (1976) 1165.
\bibitem{millis2}A.J.~Millis: Phys. Rev. B {\bf 48} (1993) 7183.
\bibitem{lahall}M.F.~Hundley, A.~Malinowski, P.G.~Pagliuso, J.L.~Sarrao and J.D.~Thompson: Phys. Rev. B. {\bf 70} (2004) 035113.
\bibitem{ziman}J.~M.~Ziman: {\it Electrons and Phonons} (1960).
\bibitem{QCP1} J.~Paglione, M. A.~Tanatar, D. G.~Hawthorn, E.~Boaknin, R.~W.~Hill, F.~Ronning, M.~Sutherland, L.~Taillefer, C.~Petrovic, P.~C.~Canfield: Phys.~Rev.~Lett. {\bf 91} (2003) 246405.
\bibitem{ebihara}T.~Ebihara, N.~Harrison, M.~Jaime, S.~Uji and J.C.~Lashley: Phys. Rev. Lett. {\bf 93} (2004) 246401.
\bibitem{kontani-RV} H.~Kontani and K.~Yamada: J. Phy. Soc. Jpn. {\bf 74} (2005) 155.
\bibitem{Yamada-textbook} K.~Yamada:{\it Electron Correlation in Metals} (Cambridge Univ. Press, 2004).
\bibitem{comment}According to the self consistent renormalization theory, $\xi_{AF}^2\propto T^{-3/2}$ just on the AF QCP in 3D systems. However, $\xi_{AF}^2$ exhibits a Curie-Weiss behavior for a wide range of temperature near the AF QCP.
\bibitem{ong3}N.P.~Ong: Phys. Rev. B {\bf 43} (1991) 193.
\bibitem{kontani-MR}H.~Kontani: J. Phys. Soc. Jpn. {\bf 70} (2001) 1873.
\bibitem{onari}S.~Onari, H.~Kontani and Y.~Tanaka: Phys. Rev. B {\bf 73} (2006).\bibitem{katayama}K.~Katayama, T.~Nagai, H.~Taniguchi, K.~Satoh, N.~Tajima and R.~Kato, to be published in J. Low Temp. Phys. (2006).
\bibitem{sushko}Y.V.~Sushko, N.Shirakawa, K.~Murata, Y.~Kubo, N.D.~Kushch, and E.B.~Yagubskii: Synth. Met. {\bf 85} (1997) 1541.
\bibitem{kontani-kappa} H.~Kontani and H.~Kino: Phys. Rev. B {\bf 63} (2001) 134524.
\bibitem{satom}S.~Yoshii, K.~Murata and M.~Sato?FJ. phys. Soc. Jpn. {\bf 69} (2000) 17.
\bibitem{VO}T.F.~Rosenbaum, A.~Husmann, S.A.~Carter and J.M~Honig: Phys. Rev. B {\bf 57} (1998) R13997.
\bibitem{drew1}J.~Cerne, M.~Grayson, D.C.~Schmadel, G.S.~Jenkins, H.D.~Drew, R.~Hughes, A.~Dabkowski, J.S.~Preston, and P.-J.~Kung: Phys. Rev. Lett. {\bf 84} (2000) 3418.
\bibitem{drew2} M.~Grayson, L.B.~Rigal, D.C.~Schmadel, H.D.~Drew, and P.-J.~Kung: Phys. Rev. Lett. {\bf 89} (2002) 037003. 
\bibitem{kontani-ACHall} H.~Kontani: J. Phys. Soc. Jpn. {\bf 75} (2006) 013703.
\bibitem{kawasaki}S.~Kawasaki, M.~Yashima, T.~Mito, Y.~Kawasaki, G.-q.~Zheng, Y.~Kitaoka, D.~Aoki, Y.~Haga and Y.~Onuki: J. Phys.: Condens. Matter {\bf 17} (2005) S889.
\bibitem{park}T.~Park, F.~Ronning, H.Q.~Yuan, M.B.~Salamon, R.~Movshovich, J.L.~Sarrao, and J.D.~Thompson, {\it et al.}: Nature, 440 (2006) 65.
\bibitem{paschen}S.~Paschen, T.~Luhmann, S.~Wirth, P.~Gegenwart, O.~Trovarelli, C.~Geibel, F.~Steglich, P.~Coleman, Q.~Si: Nature {\bf 432} (2004) 881.
\bibitem{tsujii} N.~Tsujii, H.~Kontani and K.~Yoshimura: Phys. Rev. Lett. {\bf 94} (2005) 057201. 
\bibitem{maehira} T.~Maehira, T.~Hotta, K.~Ueda and A.~Hasegawa: J. Phys. Soc. Jpn. {\bf 72} (2003) 854.
\bibitem{kontani-N} H.~Kontani: Phys. Rev. Lett. {\bf 89} (2003) 237003. 

\end{thebibliography}
\end{document}